\documentclass[acmlarge]{acmart}
\usepackage{xcolor} % Include the xcolor package

\AtBeginDocument{%
  \providecommand\BibTeX{{%
    \normalfont B\kern-0.5em{\scshape i\kern-0.25em b}\kern-0.8em\TeX}}}

\copyrightyear{2026}
\setcopyright{cc}
\setcctype{by-nc-nd}
\acmJournal{IMWUT}
\acmYear{2026} \acmVolume{10} \acmNumber{3} \acmArticle{101}
\acmMonth{9} \acmDOI{10.1145/3831630}

\usepackage{xac}
\usepackage{booktabs}
\usepackage{multirow}
\usepackage{array}
\usepackage{makecell}
\usepackage{caption}
\usepackage{siunitx}
\usepackage{placeins}
\usepackage{balance}

\usepackage{textgreek}
\usepackage{scalerel}
\usepackage[table,xcdraw]{xcolor}

\makeatletter
\def\@seccntformat#1{\@ifundefined{#1@cntformat}%
   {\csname the#1\endcsname\quad}  % default
   {\csname #1@cntformat\endcsname}% enable individual control
}
\let\oldappendix\appendix %% save current definition of \appendix
\renewcommand\appendix{%
    \oldappendix
    \newcommand{\section@cntformat}{\MakeUppercase{\appendixname~\thesection}\quad}
}

\def\@titlecasesecfont{\sffamily\large\section@raggedright}
\def\@secfont{\@titlecasesecfont\MakeUppercase}
\makeatother

\makeatletter
\let\old@thebibliography\thebibliography
\renewcommand{\thebibliography}[1]{%
  \let\@secfont\@titlecasesecfont
  \old@thebibliography{#1}%
  \interlinepenalty=10000\relax
}
\makeatother

\begin{document}

\title{AppetiteCheck: Feasibility of Momentary Vagus Nerve Stimulation as an Implicit Intervention for Eating Behavior}

\renewcommand{\shortauthors}{Gemicioglu et al.} 

\author{Tan Gemicioglu}
\orcid{0000-0001-9324-4431}
\affiliation{%
  \institution{Cornell Tech}
  \city{New York City}
  \state{NY}
  \country{USA}}
\email{tg399@cornell.edu}

\author{Jas Brooks}
\orcid{0000-0002-6142-0346}
\affiliation{%
  \institution{MIT CSAIL}
  % \city{}
  % \country{}
  \city{Cambridge}
  \state{MA}
  \country{USA}
}
\email{jasb@mit.edu}

\author{Pedro Lopes}
\orcid{0000-0001-6527-7084}
\affiliation{%
  \institution{University of Chicago}
  \city{Chicago}
  \state{IL}
  \country{USA}}
\email{pedrolopes@cs.uchicago.edu}

\author{Tanzeem Choudhury}
\orcid{0000-0002-5952-4955}
\affiliation{%
  \institution{Cornell Tech}
  \city{New York City}
  \state{NY}
  \country{USA}}
\email{tkc28@cornell.edu}

\renewcommand{\shortauthors}{Gemicioglu et al.}

\begin{abstract}
  Overeating and emotional eating are common health issues that affect people even without an eating disorder. The vagus nerve plays a critical role in the gut-brain axis, and implanted vagus nerve stimulators have been associated with reduced appetite. In this paper, we propose transcutaneous cervical vagus nerve stimulation (tcVNS) as a ubiquitous system to provide immediate, low-effort intervention during an eating episode. In a study with 24 participants, we evaluated a mobile, handheld tcVNS device during a single episode of distracted snacking. We found that participants ate 9.6\% less and 23.6\% more slowly during vagus nerve stimulation than during sham stimulation. Post-snacking satiety was the same in both conditions, while heart rate was lower during vagus nerve stimulation. The stimulation was described as subtle and barely noticeable. Overall, these results provide evidence for the feasibility of non-invasive vagus nerve stimulation as a low-attention intervention for managing eating behavior -- one that can be packaged inside ubiquitous interactive systems. As such, we extend the design space of implicit interfaces toward physiological intervention, motivating future ubiquitous systems that pair eating-related sensing with low-attention interventions.

\end{abstract}



% A "teaser" image appears between the author and affiliation
% information and the body of the document, and typically spans the
% page.

\begin{teaserfigure}
  \centering
  \includegraphics[width=\textwidth]{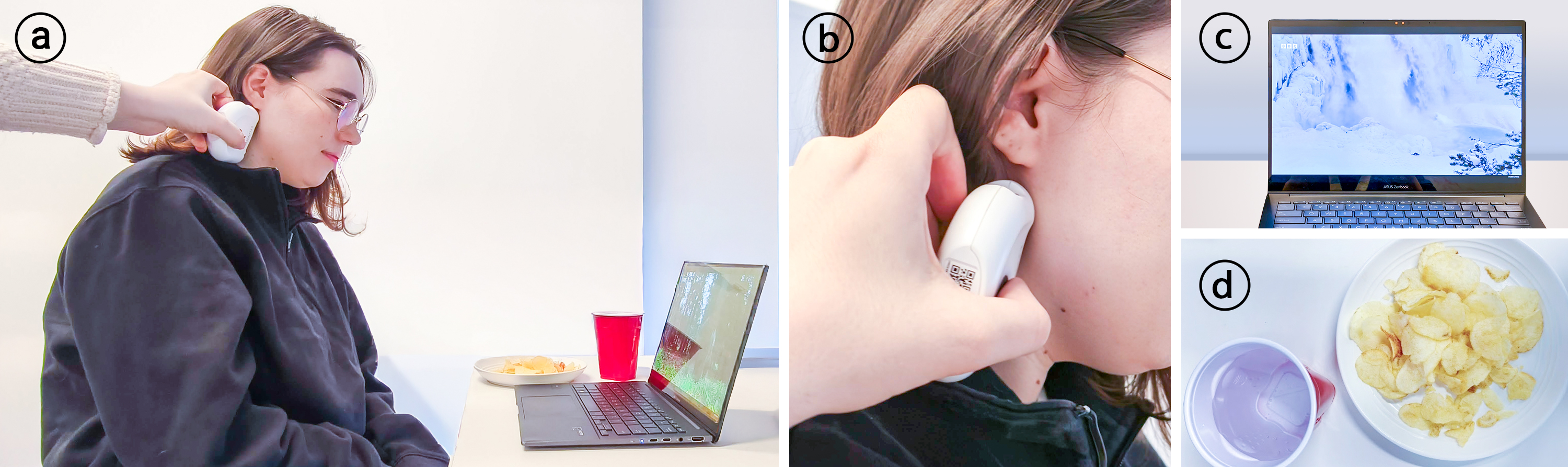}
  \caption{\textbf{AppetiteCheck: a physiological intervention for uncontrolled eating.} AppetiteCheck intervenes during distracted eating by targeting satiety-related physiological processes, without requiring perceptible feedback or sustained cognitive effort. (a) Participants snack while watching a self-selected video, representing a common distracted eating scenario. (b) The AppetiteCheck device is a ubiquitous interface that can deliver transcutaneous cervical vagus nerve stimulation (tcVNS) to the right side of the neck. Example (c) video content viewed and (d) snack (chips) and water provided in study.
  }
  \Description{Composite figure showing the AppetiteCheck study setup. One panel shows a participant snacking while watching a laptop as the stimulator is held to the neck. A close-up panel shows the stimulator on the right side of the neck. Additional panels show example video content on a laptop and a plate of chips with water.}
  \label{fig:teaser}
\end{teaserfigure}
% Fig 1: system in study configuration, video up, plate down
% Table 1: eating interventions and their effectiveness
% Fig 2: system zoomed in, vagus nerve path and location
% Table 2: all measures and questions
% Fig 3: Procedure flowchart
% Fig 4: Eating amount and rate by condition
% Table 3: statistical significance results
% Fig 5: Satiety before and after by condition
% Fig 6: Perceived effectiveness, control, embodiment by condition
% Fig 7: RMSSD-HRV before and during study, by condition, HF-HRV during study by condition

% \externaldocument{08_appendix}

\maketitle

\section{Introduction}
Eating has been a key research area in ubiquitous computing and human-computer interaction (HCI), as diet, nutrition, and eating behavior are foundational to a healthy lifestyle. 

In this investigation, we focus on exploring the opportunity for ubiquitous systems to act at the moment of eating, in a way that does not require active effort from the end-user, but instead aims to engage the body's physiological pathways to influence eating behavior. We focus on a physiological approach since it may provide a novel low-effort pathway for ubiquitous health interventions, one that can fit in eating situations where users are not mindful of their eating habits (e.g., distracted snacking while consuming media). 

In fact, overeating and emotional eating are growing problems in the US, and lead to increased risk in a number of conditions, including coronary heart disease, stroke, and Type II diabetes~\cite{dakanalisAssociationEmotionalEating2023}. Many people engage in these behaviors at a subclinical level, despite not having a diagnosed eating disorder~\cite{peschelSubclinicalPatternsDisordered2024}. As aforementioned, these behaviors are often exacerbated by situations where users are not mindful of their eating habits, such as snacking at night or while distracted by digital media. 

Prior work aiming at interactive interventions in this space has examined food journaling~\cite{cordeiroBarriersNegativeNudges2015,jungFoundationsSystematicEvaluation2020} and photo-based dietary tracking~\cite{chungIdentifyingPlanningIndividualized2019}, demonstrating the value of understanding people's eating habits and dietary intake. However, these systems have also shown the cognitive burden and harmful nudges of actively tracking eating behavior~\cite{cordeiroBarriersNegativeNudges2015}. Mobile and wearable sensing systems have sought to reduce this burden by automatically detecting eating episodes~\cite{thomazPracticalApproachRecognizing2015,bedriEarBitUsingWearable2017a,biAuracleDetectingEating2018,zhangNeckSenseMultisensor2020,shahiDetectingEatingSocial2023,liangDetectingEatingEvents2026a}, chewing~\cite{chengActivityRecognitionNutrition2013,bedriEarBitUsingWearable2017a,biAuracleDetectingEating2018,zhangNeckSenseMultisensor2020}, drinking~\cite{mahmudMunchSonicTrackingFinegrained2024,shahiDetectingEatingSocial2023}, hand-to-mouth gestures~\cite{thomazPracticalApproachRecognizing2015,mahmudMunchSonicTrackingFinegrained2024,liangDetectingEatingEvents2026a}, and fine-grained dietary actions~\cite{chengActivityRecognitionNutrition2013,mahmudMunchSonicTrackingFinegrained2024} using smartwatches, earables, neck-worn sensors, rings, cameras, and eyeglasses. Beyond tracking, mobile health systems have explored diet management and eating interventions, including systems that support reflection on dietary choices~\cite{chungIdentifyingPlanningIndividualized2019,nakamura_eat2pic_2023}, encourage balanced diets~\cite{nakamura_eat2pic_2023}, or regulate eating pace~\cite{chen_sspoon_2022,liu_hicclip_2024,chenViFeedPromotingSlow2025}.

Taken together, there has been a progression from dietary self-report to automated sensing and eventually towards behavioral closed-loop systems to connect eating tracking with timely interventions~\cite{fang_review_2025}. However, existing intervention strategies act through information, reflection, perception, or deliberate behavior change: they ask users to log, notice, interpret, slow down, or respond to a cue. These approaches can be limited during eating episodes that are habitual, distracted, and affectively driven. In ubiquitous computing, mindless interventions have explored perceptual modalities to deliver behavioral interventions while remaining at the periphery of attention \cite{adams_mindless_2015,costaBoostMeUpImprovingCognitive2019a, chenViFeedPromotingSlow2025}. AppetiteCheck explores a complementary interactive approach: momentary physiological modulation through non-invasive vagus nerve stimulation.

% In ubiquitous computing, several methods for mobile and wearable eating detection have been developed to capture eating events alongside their context~\cite{thomazPracticalApproachRecognizing2015,bedriEarBitUsingWearable2017a,mahmudMunchSonicTrackingFinegrained2024}. Characterizing nutrition and eating behavior has been a key focus area in recent activity recognition systems, integrating these advancements into everyday devices \cite{}. For these sensing systems to promote healthy behavior, they are often coupled with a health intervention \cite{}.

% However, current eating interventions cannot affect eating with the temporal specificity and unobtrusiveness of wearable sensing systems. Eating interventions are limited to slow-acting pharmacological approaches that affect all eating behavior, or high-effort digital tools that interfere with eating to encourage mindful, self-regulatory eating practices. There remains a critical gap for a low-effort, momentary intervention that can modulate the physiological drivers of appetite without requiring the user to exert intense willpower or pause their primary activity.

% Vagus nerve stimulation
The vagus nerve serves as the primary bidirectional communication pathway of the gut-brain axis, regulating satiety signals and the balance of the autonomic nervous system (ANS). Past research has shown that vagus nerve stimulation (VNS) via surgical implants can reduce weight and food intake in both animal models~\cite{gil_electrical_2011, yao_effective_2018} and humans~\cite{pardo_weight_2007}. Yet, these implants are not without health complications and need to be replaced or removed with repeated surgical procedures. 

% Moreover, due to the strength of the stimulation and its proximity to other nerve pathways in the neck, there are common side effects, such as changes in voice or breathing \cite{}.

Transcutaneous cervical vagus nerve stimulation (tcVNS) offers a non-invasive alternative in which a mild electrical current is delivered to the side of the neck. Recent work has shown that even non-invasive, transcutaneous vagus nerve stimulation can increase stomach-brain coupling~\cite{muller_vagus_2022}. Moreover, the effects of VNS can take place rapidly, making it suitable as a momentary intervention~\cite{rodenkirch_rapid_2022,li_immediate_2023}.

In this paper, we evaluate the feasibility of tcVNS as a momentary, low-attention intervention during eating. We introduce AppetiteCheck, an approach that applies tcVNS during an eating episode to examine whether short-term vagus nerve stimulation is associated with measurable differences in eating behavior, satiety, and physiological activity. We examine tcVNS as a potential intervention modality for future closed-loop eating systems, while exploring it here in a constrained laboratory setting using an experimenter-operated handheld device.

To evaluate AppetiteCheck, we conducted a within-subject randomized controlled trial with 24 participants who were asked to eat snacks while watching videos in a controlled lab setting. While eating, participants received vagus nerve stimulation on their neck or sham stimulation on their shoulder, as shown in Figure \ref{fig:teaser}. The study aimed to answer the following research questions:
\begin{itemize}
    \item[\textbf{RQ1:}] How does momentary tcVNS affect the amount of food eaten and eating rate in a single eating episode?
    \item[\textbf{RQ2:}] How is satiety after an eating episode affected by momentary tcVNS?
    \item[\textbf{RQ3:}] What is the effect of momentary tcVNS on physiological activity?
\end{itemize}

Our findings indicate that participants ate 9.6\% less and 23.6\% more slowly during tcVNS than during sham stimulation. Despite eating less, participants receiving tcVNS still felt satiated after snacking. Participants had lower heart rates during tcVNS, suggesting a physiological response. Participants described the stimulation as subtle and barely noticeable. Overall, these results provide initial evidence that momentary tcVNS is associated with changes in eating behavior during a controlled snacking task.

Through our evaluation of AppetiteCheck, we make three main contributions to the field of behavioral interventions and ubiquitous computing:
\begin{itemize}
    \item We conducted, to the best of our knowledge, the first evaluation of tcVNS during eating.
    \item We outline a path toward closed-loop ubiquitous devices that provide implicit interventions on eating behavior by showing that momentary tcVNS was associated with lower food intake during distracted snacking.
    \item We extend the design space of mindless computing by demonstrating the feasibility of a device-based physiological intervention for eating as an alternative to sensory cues in a controlled lab setting.
\end{itemize}

% Our findings

% Through a randomized controlled trial with 24 participants, we demonstrate that tcVNS is a viable, non-intrusive method for reducing food intake and eating speed without negatively impacting the user's sense of satiety.

\section{Related Work}

\subsection{Vagus Nerve Stimulation}

%implants, tavns, tcvns
Vagus nerve stimulation (VNS) involves the electrical stimulation of the vagus nerve by an electronic device. It emerged clinically through implantable nerve stimulators for epilepsy~\cite{zabaraInhibitionExperimentalSeizures1992} and was later extended to severe depression~\cite{sackeimVagusNerveStimulation2001}, but surgical implantation limits its practicality for milder or everyday-use contexts. Transcutaneous approaches reduce this barrier: transcutaneous auricular VNS (taVNS) stimulates the cymba concha of the ear and has been studied for migraines~\cite{straubeTreatmentChronicMigraine2015}, post-traumatic stress disorder (PTSD)~\cite{powersVagusNerveStimulation2025}, and stroke rehabilitation~\cite{dawsonVagusNerveStimulation2020}. Transcutaneous cervical VNS (tcVNS) instead stimulates the neck, providing easier access than the ear and requiring less precision in electrode positioning, making it more compatible with wearable or self-administered devices.

VNS is typically delivered as \textit{chronic}, \textit{episodic}, or \textit{momentary} stimulation. Chronic stimulation operates continuously, such as 30 seconds every 5 minutes for treatment-resistant depression or epilepsy~\cite{sackeimVagusNerveStimulation2001, elliottEfficacyVagusNerve2011}. Episodic stimulation is delivered in discrete therapy sessions, as in many scientific studies~\cite{teckentrup_non-invasive_2020} and migraine treatments~\cite{straubeTreatmentChronicMigraine2015}. Momentary stimulation instead occurs in parallel with a task, leveraging rapid effects of VNS to enhance plasticity during rehabilitation~\cite{ruizEffectiveDeliveryVagus2023, dawsonVagusNerveStimulation2020} and sensory processing~\cite{rodenkirch_rapid_2022, jigo_transcutaneous_2024}. AppetiteCheck investigates whether this momentary paradigm can extend to eating behavior and appetite.

VNS has also been linked to appetite and weight management. Patients with obesity who received implanted VNS for other conditions, such as depression, were observed to lose weight~\cite{pardo_weight_2007}. Later, clinical trials evaluated abdominal vagal blockade, which uses surgically implanted devices to block hunger signals with high-frequency pulses~\cite{camilleriIntraabdominalVagalBlocking2008a}; longitudinal studies found weight reductions in patients with obesity~\cite{apovian_two-year_2017, shikora_sustained_2015}.

Non-invasive VNS can also affect gut-brain pathways: taVNS has been shown to reduce gastric frequency~\cite{teckentrup_non-invasive_2020} and increase stomach-brain coupling through a vagal afferent pathway~\cite{muller_vagus_2022}. The closest study to AppetiteCheck evaluated taVNS for food intake and found no effect~\cite{obstEffectShortTermTranscutaneous2020a}; however, it used episodic taVNS before eating rather than momentary stimulation during eating, and used a very low current level (0.6mA) compared to the tcVNS approach evaluated here.

%% This could be cut down by 100 words (lots of enumeration)
\subsection{Implicit Health Interventions}
Implicit health interventions aim to support behavior change without requiring users to stop their primary activity or continuously attend to the intervention. This goal builds on implicit and peripheral interaction, where systems respond to users' activity, context, or inferred intent without requiring explicit commands or sustained focal attention~\cite{schmidtImplicitHumanComputer2000,juDesignImplicitInteractions2008,juDesignImplicitInteractions2015,bakkerPeripheralInteractionCharacteristics2015}. In health behavior change, this framing appears in JITAIs, which adapt support to users' state and context~\cite{nahumshaniJustintimeAdaptiveInterventions2018}, and in \textit{mindless computing}, which leverages automatic responses and implicit processes rather than relying primarily on attention, motivation, or conscious effort~\cite{adams_mindless_2015}.

Prior HCI and UbiComp work has explored several perceptual routes for implicit health interventions. Some systems alter users' perception of their own behavior or bodily state: changing how people hear their own voice during interpersonal conflict can reduce anxiety or increase perceived power~\cite{costaRegulatingFeelings2018}, while heartbeat-like tactile feedback has been used to regulate emotion, stress, or physiology during ongoing tasks~\cite{azevedoCalmingEffectNew2017,costaBoostMeUpImprovingCognitive2019a,choiAmbienBeatWristworn2020}. Others use sensory stimulation as an affective or regulatory cue, including wearable affective touch~\cite{zhaoAffectiveTouchImmediate2023a}, warm or pressure-based haptic materials~\cite{papadopoulouAffectiveSleeve2019}, olfactory interfaces~\cite{amoresEssenceOlfactory2017,amoresBioEssenceWearableOlfactory2018}, and breathing guidance through haptic, audio, or airflow cues~\cite{paredesJustBreathe2018,gemiciogluBreathePulsePeripheralGuided2024b}.

Across these examples, implicit health interventions typically operate by changing what users perceive: their voice, apparent heartbeat, tactile environment, scent, airflow, sound, or other sensory feedback. \textit{AppetiteCheck} extends this framing from perceptual and contextual interaction to physiological actuation: the intervention remains peripheral to the eating activity, but its mechanism is neuromodulation rather than a cue that users must notice, interpret, follow, or entrain to during an ongoing eating episode.

\subsection{Eating Management \& Eating Interventions}

Ubiquitous computing and HCI researchers have explored eating-related systems for several goals, including nutrition tracking~\cite{chengActivityRecognitionNutrition2013,jungFoundationsSystematicEvaluation2020}, dietary self-monitoring~\cite{cordeiroBarriersNegativeNudges2015,chungIdentifyingPlanningIndividualized2019}, eating detection~\cite{bedriEarBitUsingWearable2017a,biAuracleDetectingEating2018,zhangNeckSenseMultisensor2020,shahiDetectingEatingSocial2023,mahmudMunchSonicTrackingFinegrained2024,liangDetectingEatingEvents2026a}, diet management~\cite{chungIdentifyingPlanningIndividualized2019,grahamIntegratingUserCentered2021}, and eating behavior modification~\cite{nakamura_eat2pic_2023,chen_sspoon_2022,liu_hicclip_2024,chenViFeedPromotingSlow2025}. For the physical action of eating, prior systems have also manipulated food perception and dining experience~\cite{linFoodFabCreating2020,ranasingheAugmentedFlavours2019}, examined screen-based dining and mindful eating~\cite{khotUnderstandingScreenbasedDining2022,khotMindEatNavigatingScreenCentric2025,chenViFeedPromotingSlow2025}, or attempted to change eating pace independently of overall food intake~\cite{hermansEffectRealtimeVibrotactile2017a,melansonEatingPaceInstruction2023}.

Prior work has also connected eating sensing with timely or closed-loop intervention. For example, Rahman et al. predicted ``about-to-eat'' moments from wearable sensor streams to support just-in-time intervention~\cite{rahmanPredictingAbouttoEat2016}, eat2pic reflects sensed eating behavior through immediate and delayed visual feedback~\cite{nakamura_eat2pic_2023}, and Earinter uses commodity earbuds for closed-loop eating-pace regulation through just-in-time audio feedback~\cite{fangEarinterClosedLoopSystem2026}. As eating detection becomes more robust, closed-loop systems need intervention mechanisms that can be delivered at the moment of eating without adding substantial attentional or behavioral demands.

\textit{AppetiteCheck} is an intake-focused eating intervention: a system that aims to slow, reduce, or control food intake. As similar behavioral outcomes can arise from different mechanisms, we compare these systems by how they act during eating rather than by device form: sensory interventions alter the eating experience, cognitive interventions recruit attention or self-regulation, and physiological interventions target appetite or satiety pathways. Table~\ref{tab:intervention_comparison} summarizes examples from this space.

\textbf{Sensory interventions.} Sensory interventions influence consumption by changing how food or eating is perceived, often without explicit instruction. SSpoon alters utensil shape to reduce bite size, encouraging slower eating and modest reductions in food intake~\cite{chen_sspoon_2022}; vibrotactile forks prompt slower eating or increased awareness of eating pace~\cite{hermansEffectRealtimeVibrotactile2017a}; Hicclip sonically augments eating sounds, reducing eating duration~\cite{liu_hicclip_2024}; and ViFeed adjusts video playback speed during meals to influence eating rate~\cite{chenViFeedPromotingSlow2025}. These systems can operate at the periphery of attention, but their effectiveness still depends on users perceiving and integrating the sensory cue.

\textbf{Cognitive interventions.} Cognitive interventions explicitly engage attention, motivation, or self-control through smartphone reminders~\cite{whitelockSmartphoneBasedAttentive2019}, verbal pacing guidance~\cite{melansonEatingPaceInstruction2023}, or attentive and mindful eating strategies~\cite{robinson_eating_2013,higgs_manipulations_2015,seguias_effect_2018}. These approaches can change eating behavior, but typically require sustained effort and compliance. Reliance on executive control can limit long-term adoption, particularly when motivation and cognitive bandwidth fluctuate~\cite{hallExecutiveControlResources2014,adams_mindless_2015}.

\textbf{Physiological interventions.} Physiological interventions instead target internal appetite and satiety pathways. GLP-1 receptor agonists, for example, can substantially reduce food intake and body weight by modulating gut-brain signaling~\cite{gutzwillerGlucagonlikePeptide1Potent1999, gibbonsEffectsOralSemaglutide2021}. These interventions are largely passive from the user's perspective but introduce tradeoffs related to side effects, invasiveness, cost, and long-term adherence.

Prior work has examined implanted VNS and vagal blockade for appetite and weight management~\cite{pardo_weight_2007,camilleriIntraabdominalVagalBlocking2008a,apovian_two-year_2017,shikora_sustained_2015}, but momentary, non-invasive physiological actuation during eating remains comparatively underexplored as an intervention mechanism. \textit{AppetiteCheck} explores this space by applying momentary tcVNS during eating and measuring its effects on food intake, eating rate, appetite, and physiological signals. Unlike sensory or cognitive systems, \textit{AppetiteCheck} does not aim to alter how food is perceived or require users to monitor their behavior. Instead, it evaluates tcVNS as a candidate physiological intervention that may operate alongside eating without demanding sustained attention.

This comparison does not collapse the different contexts in which intake-focused interventions are used: systems for obesity, binge eating disorder, anorexia nervosa, or general dietary support can share behavioral targets while raising different clinical and safety concerns~\cite{grahamIntegratingUserCentered2021,grahamApplyingUserCentered2021,hilbertSmartphoneSupportedCognitive2025,schleglPotentialTechnologyBased2015}. We address this further in Section \ref{wearable-discussion}.

% Mindful eating approaches, apps
% \cite{nakamura_eat2pic_2023} \cite{xie_e-reminder_2023}

% Devices and mindless interventions
% \cite{adams_mindless_2015} 

\renewcommand{\arraystretch}{1.5}
\begin{table}[]
\caption{\textbf{Comparison of eating interventions and reported behavioral effects.}
Prior work and AppetiteCheck are compared by modality, mechanism, user engagement, sample size, and reported changes in food intake and eating rate. AppetiteCheck differs from sensory and cognitive interventions by using physiological modulation while remaining at the periphery of attention. In this controlled snacking task, AppetiteCheck produced a food intake reduction comparable to several non-pharmacological interventions and a low-dose pharmacological intervention.}
\label{tab:intervention_comparison}
\resizebox{\textwidth}{!}{%
\begin{tabular}{ccccccc}
\hline
\rowcolor[HTML]{EFEFEF} 
\textit{Title}           & \textit{Modality}            & \textit{Mechanism}     & Active?          & \textit{\begin{tabular}[c]{@{}c@{}}Sample\\ Size\end{tabular}} & \textit{\begin{tabular}[c]{@{}c@{}}Food Intake\\ Reduction (\%)\end{tabular}} & \textit{\begin{tabular}[c]{@{}c@{}}Eating Rate\\ Reduction (\%)\end{tabular}} \\ \hline
\textbf{AppetiteCheck}   & \textbf{Device (electrical)} & \textbf{Physiological} & \textbf{Passive} & \textbf{24}                                                    & \textbf{9.6\%}                                                                & \textbf{23.6\%}                                                               \\
\rowcolor[HTML]{EFEFEF} 
SSpoon (2022) \cite{chen_sspoon_2022}            & Device (shape)               & Sensory                & Passive          & 42                                                             & 4.5\%                                                                         & 14.9\%                                                                        \\
Hermans et al. (2017) \cite{hermansEffectRealtimeVibrotactile2017a}    & Device (vibration)           & Sensory                & Active           & 114                                                            & -1.6\%                                                                        & 13.8\%                                                                        \\
\rowcolor[HTML]{EFEFEF} 
ViFeed (2025) \cite{chenViFeedPromotingSlow2025}           & Digital (video speed)        & Sensory                & Conditional      & 24                                                             & 9.6\%                                                                         & 15.3\%                                                                        \\
Whitelock et al. (2019) \cite{whitelockSmartphoneBasedAttentive2019}  & Digital (reminders)          & Cognitive              & Active           & 107                                                            & 8.4\%                                                                         & -                                                                             \\
\rowcolor[HTML]{EFEFEF} 
Melanson et al. (2023) \cite{melansonEatingPaceInstruction2023}  & Education (verbal)           & Cognitive              & Active           & 50                                                             & -1.7\%                                                                        & 37.4\%                                                                        \\
Gutzwiller et al. (1999) \cite{gutzwillerGlucagonlikePeptide1Potent1999} & Drug (IV)                    & Physiological          & Passive          & 16                                                             & \begin{tabular}[c]{@{}c@{}}Low Dose: 9.5\%\\ High Dose: 34.2\%\end{tabular}   & \begin{tabular}[c]{@{}c@{}}Low Dose: 1.8\%\\ High Dose: 17.1\%\end{tabular}   \\
\rowcolor[HTML]{EFEFEF} 
Gibbons et al. (2021) \cite{gibbonsEffectsOralSemaglutide2021}    & Drug (Oral)                  & Physiological          & Passive          & 13                                                             & 38.9\%                                                                        & -                                                                            
\end{tabular}
}
\end{table}

\section{Intervention Design}

% Pedro: Would appreicate help here, as you did in BreathePulse

\subsection{System}

\label{system-design}

To deliver tcVNS, we chose Truvaga Plus~\cite{TruvagaHome}, an off-the-shelf, FDA-cleared handheld vagus nerve stimulator shown in Figure~\ref{fig:mechanism} for replicability, safety, and compliance with our institutional IRB. The Truvaga Plus mimics the hardware of an FDA-approved device, gammaCore, built by the same company and in clinical use for a variety of conditions, including migraines and cluster headaches~\cite{mwamburiReviewNoninvasiveVagus2017}. The Truvaga Plus stimulator delivers 5 cycles of 5kHz biphasic sinusoidal pulses at 25Hz, with a pulse width of 1ms. The stimulation strength can be adjusted using a phone application, with 10 levels corresponding to different levels of current, up to 60mA.

Although we initially designed a wearable prototype, consisting of a pocket-sized stimulator and gel electrode patches modeled after past work on electrical muscle stimulation in HCI \cite{lopesAffordanceAllowingObjects2015,lopesImmensePowerTiny2017}, we ultimately prioritized safety and physiological validity for our feasibility testing by using an FDA-cleared device. We discuss plans and design challenges for a compact and comfortable wearable vagus nerve stimulator in Section \ref{wearable-discussion}.

\subsection{Intervention Approach} \label{approach}

% I should add more info here

\begin{figure}
    \centering
    \includegraphics[width=\textwidth]{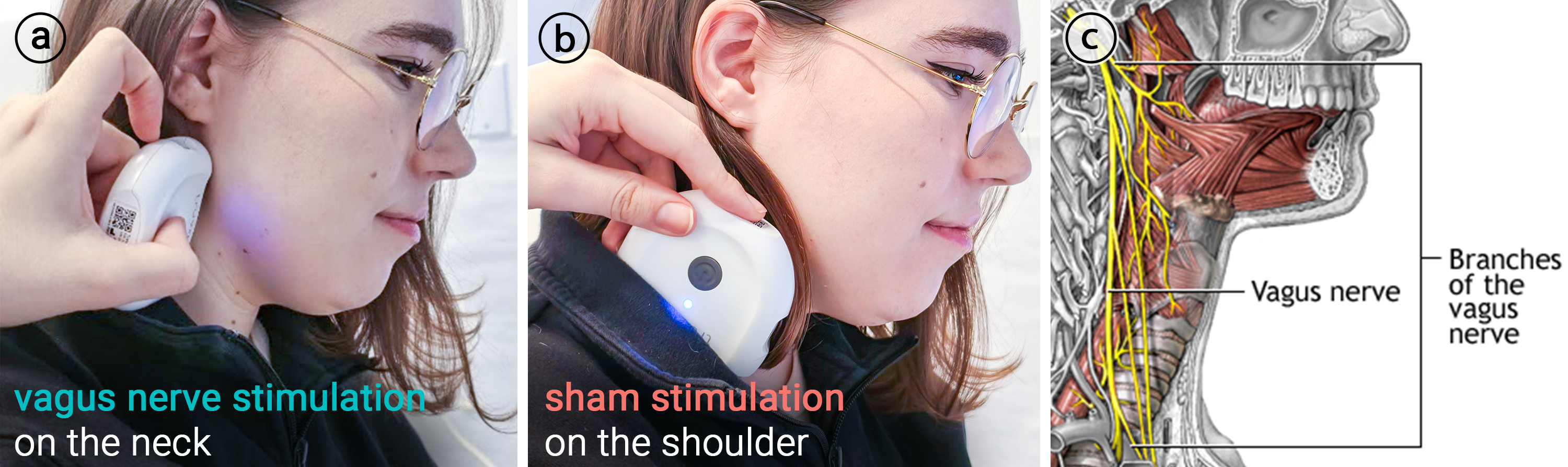}
    \caption{\textbf{Stimulation conditions and anatomical target.} 
    (a) Active tcVNS delivered on the right side of the neck. (b) Sham stimulation on the shoulder to control for device contact and handling. (c) Anatomical illustration showing the cervical course and branches of the vagus nerve targeted by tcVNS.}
    \Description{Three-panel figure comparing stimulation locations. The first panel shows active tcVNS delivered on the right side of the neck with the handheld stimulator. The second panel shows sham stimulation delivered on the shoulder. The third panel is an anatomical illustration labeling the vagus nerve and branches of the vagus nerve in the neck.}
    \label{fig:mechanism}
\end{figure}

The stimulation for AppetiteCheck was delivered by holding the handheld stimulator against the side of the neck. As shown in Figure \ref{fig:mechanism}, this location is one of the areas where the vagus nerve comes closest to the surface of the skin, allowing electrical pulses to propagate to the nerve. In the case of sham stimulation, strength calibration and stimulation were both done on the top of the shoulder instead. The stimulation followed a cycle of 2 minutes on and 2 minutes off, repeated up to 5 times, for a total of 20 minutes, during which stimulation was active for 10 minutes. The on/off cycle was based on protocols for other applications of the gammaCore \& Truvaga devices. While the device was experimenter-operated to ensure consistent placement and participant safety, this setup allowed participants to experience stimulation without actively handling the device, monitoring timing, or interrupting the snacking task.

The key ideas behind AppetiteCheck's VNS approach were to deliver stimulation with minimal perceptible sensation and to intervene in the moment of eating. Typically, nerve stimulation systems are calibrated by starting at a low current level and gradually adjusting to the maximum level deemed ``tolerable'' and not painful by participants~\cite{jigo_transcutaneous_2024}. As a result, these systems may result in physically uncomfortable tingling sensations. AppetiteCheck aimed to minimize discomfort to ensure the system could function effectively in the background without drawing the user's attention. AppetiteCheck was calibrated for each participant by adjusting the strength only up to the level at which participants reported a sensation on their skin. The calibration was performed on the same location where they would be receiving stimulation during the session, using a mobile app controlled by the experimenter, which could incrementally change stimulation strength.

To evaluate ``momentary'' stimulation, stimulation began concurrently with the eating section of the experiment. While the participant was focused on the video and snacking, the experimenter used a timer to deliver VNS or sham stimulation for 2 minutes, followed by a 2-minute rest period. The stimulator was only held against the target location for the stimulation duration and placed on the table at all other times. To avoid exceeding 10 minutes of stimulation, no more stimulation was delivered after the first 20 minutes of eating.

To ensure effective delivery, Signagel electrode gel was applied to the device's electrodes prior to stimulation. Stimulation was always delivered on the right side of the neck for consistency with past studies investigating the gut-brain connection of the vagus nerve \cite{muller_vagus_2022}.

\section{Methods}  \label{study_design}
We conducted a controlled within-subject study to evaluate vagus nerve stimulation as a momentary intervention during distracted snacking. The study was approved by the Cornell University Institutional Review Board (IRB0148698).

\subsection{Participants}

We recruited 27 participants using mailing lists, Slack, and posters around the university campus. Eligible participants were aged 18 to 65 and had none of the following conditions: Any eating disorder, any implanted electronic device, pregnancy, vagotomy, active cancer or cancer in remission, abnormal cervical anatomy, carotid atherosclerosis, clinically significant hypertension or hypotension, aneurysms, and history of cardiac disease, arrhythmia, brain tumor, or seizure. Participants completed a screening form when they signed up for the study to verify that they met the criteria. Participants were also screened for food allergies and other dietary restrictions to ensure the snacks they ate during the study met their requirements. 

All participants provided written informed consent. Three participants were excluded after a one-hour session and compensated with a \$30 Amazon gift card: two for being unable to schedule their second study session, and one for not eating any snacks during the study. The remaining 24 participants completed both sessions of the study, for one hour each, and were compensated with a \$60 Amazon gift card. For the remainder of the paper, we only refer to the 24 participants who fully completed the study.

% Demographics
Participants with complete data were aged 22 to 53 (\(M = 28.3, SD = 6.9\)), of whom 12 were male, 10 were female, and 2 were non-binary. Ethnically, 11 participants were Asian, 10 were White, 3 were Hispanic, 2 were Middle Eastern or North African, and 1 was Black or African-American. 17 participants reported not having experienced any electrical stimulation, while 7 had experienced some form of electrical stimulation, such as transcutaneous electrical nerve stimulation or muscle stimulation, but not VNS.

%Power analysis 

An a priori power analysis was performed with G*Power version 3.1~\cite{faul2007g} to determine the minimum number of participants required to evaluate the effect of VNS. To achieve 80\% power with a medium-sized effect of 0.25 and a significance level of α = 0.05, a sample size of N = 21 is necessary for a repeated measures ANOVA and within-subjects factors. Consequently, the achieved sample size of N = 24 is sufficient for examining the study's hypotheses.

\subsection{Measures} \label{measures}

To evaluate AppetiteCheck's primary effect on eating, we quantitatively measured participants' eating behavior with physical instruments. We administered a questionnaire to collect subjective and perceptual information about the stimulation and the participant. Using a heart rate monitor, we assessed physiological responses to momentary tcVNS. Lastly, we collected qualitative feedback on the experience through open-ended questions at the end of each study session. 

%The full list of measures is summarized in Table \ref{tab:measures}.

\subsubsection{Behavioral Measures} \hfill

\textbf{Food Weight.} At the start of each study session, participants were offered a standard amount of their chosen snack, further described in Section \ref{procedure}, and weighed in grams using an electronic scale. At the end of each study session, the plate containing the snacks was weighed again, and the difference was recorded as the amount of food the participant had eaten. Because the offered amount varied by snack type, we also calculated the percentage of the offered amount each participant ate, which was used in statistical analyses of total food intake.

\textbf{Eating Duration.} Each participant was given a maximum of 30 minutes for eating. If participants finished all offered snacks or told the experimenter they were full early, the eating section of the study was stopped early. The duration of eating was measured in minutes using a timer, starting with the first instance of vagus nerve or sham stimulation. The eating duration was used to evaluate the participants' eating rate.

\textbf{Eating Rate.} By dividing the amount of food eaten by the duration of the eating episode, we also calculated participants' eating rate in grams per minute. Eating rate was analyzed as a complementary outcome because it captures changes in eating pace for both participants who stopped early and participants who did not finish the full portion, and because it is commonly reported in prior eating intervention work.

\subsubsection{Self-Reported Measures} \hfill

\textbf{Satiety.} To evaluate the effect of stimulation on satiety, we used a commonly used satiety questionnaire by Flint et al. \cite{flintReproducibilityPowerValidity2000}. The questionnaire consists of four questions: ``How hungry do you feel?'', ``How satisfied do you feel?'', ``How full do you feel?'', ``How much do you think you can eat?'' These are presented on a 7-point Likert scale and averaged, with the first and fourth questions reverse-coded. We asked participants to fill out the questionnaire both before and after eating in each session.

\textbf{Eating Impulsivity.} Questionnaires on eating traits measure uncontrolled eating behaviors at different severity levels. The mildest form is the Eating Impulsivity questionnaire, which is suitable for evaluating uncontrolled eating in healthy participants \cite{vainik_eating_2015}. The Eating Impulsivity questionnaire consists of three 7-point Likert scale questions from the NEO Personality Inventory-3's N-5 Impulsiveness scale \cite{vainikAreTraitOutcome2015}: ``It is hard for me to control my impulses.'', ``I tend to eat too much of my favorite food.'', ``Sometimes I am not able to control my appetite.'' Participants completed these questions at the start of a study session. Given our healthy participant group, we used this questionnaire to measure potential interaction effects due to uncontrolled eating. For the purpose of statistical analyses, we coded participants with average ratings above 4 as ``Impulsive'' and below 4 as ``Not Impulsive'' as a categorical variable. This approach evenly divided participants into 12 Impulsive and 12 Not Impulsive.

\textbf{Stimuli Perception.} We asked participants to rate their perception of the stimuli after each study session. The ratings were along three dimensions: strength (``How strongly did you perceive the stimulation?''), frequency (``How often did you perceive the stimulation?''), and perceived effect (``Did the stimulation affect your eating?''). Each was presented as a 7-point Likert scale.

\textbf{Agency \& Embodiment.} As a device closely coupled to physiological activity, AppetiteCheck poses interesting questions on agency and the degree to which interventional devices can integrate with the body \cite{muellerNextStepsHumanComputer2020a}. To evaluate these, we used two questions designed by Bergström et al. for evaluating interaction techniques, rephrased for the eating application \cite{bergstrom_sense_2022}: ``It felt like I was in control of my eating during the task.'' and ``It felt like the device I was using was a part of my body.'' Each was presented as a 7-point Likert scale at the end of each study session.

% Typically, agency in HCI is evaluated using statements such as ``I did it.'' However, this statement loses its meaning when applied to autonomic processes. AppetiteCheck is not designed to directly affect whether participants eat or whether they stop eating. Instead, it is designed to affect appetite, which influences their decision to stop eating. 

\subsubsection{Physiological Measures} \hfill

\textbf{Heart Rate.} During the study, each participant wore a Polar H10 heart rate monitor, which records electrocardiogram (ECG) signals at 130Hz. The heart rate monitor was located on the chest and in contact with skin throughout the study session. The mean heart rate for three time intervals -- before eating, during eating, and after eating -- was extracted from the ECG signals. Lower heart rate is generally associated with increased parasympathetic activity, but is insufficient to establish an autonomic response \cite{gordanAutonomicEndocrineControl2015}.

\textbf{Heart Rate Variability (HRV).} In addition to the heart rate, we calculated two HRV metrics to evaluate the physiological response to AppetiteCheck. Firstly, we extracted the root mean square of the interval between consecutive heartbeats (RMSSD). Higher RMSSD is associated with parasympathetic activity, while lower RMSSD is often used to express physiological stress \cite{shafferOverviewHeartRate2017}. We also calculated high-frequency (HF) band power in the frequency domain. Higher HF power is associated with vagal activity, but has a more complex relationship due to multiple factors affecting the metric. For statistical evaluation, we used the natural logarithms of the two HRV metrics, lnRMSSD and lnHF, as they are log-normally distributed \cite{shafferOverviewHeartRate2017}.

\subsubsection{Qualitative Feedback} \hfill

\textbf{Sensory Analysis.} Participants provided open-ended written feedback across three questionnaire items: ``How did you feel about the sensation from the stimulation?'', ``What was your experience like with the device?'', and ``Is there anything else you would like to add?''. To compare the tcVNS and sham conditions, responses were annotated into five dimensions: (1) \textit{sensation}, (2) \textit{intensity}, (3) \textit{temporal characteristics}, (4) \textit{likeness}, and (5) \textit{secondary impacts}.

Descriptors were normalized by collapsing semantically similar expressions while preserving perceptually meaningful distinctions. Descriptions clearly attributable to manual pressure from the experimenter were excluded, and annotations were performed by a single annotator with experience in qualitative sensory analysis.

\textbf{Form-Factor Feedback.} Form-factor comments were drawn from responses to ``What was your experience like with the device?'' and included only explicit mentions of pressure, device placement, manual holding, or the experimenter. These comments were extracted and analyzed separately from sensory descriptors.

\textbf{Adoption in Daily Life.} Participants were also asked ``Would you use this device in daily life? If so, how?'' Responses were coded to capture overall adoption stance (\textit{yes}, \textit{maybe}, \textit{no}), conditions or reasons underlying that stance, and intended use cases.

\subsection{Study Procedure} \label{procedure}

\begin{figure}
    \centering
    \includegraphics[width=\linewidth]{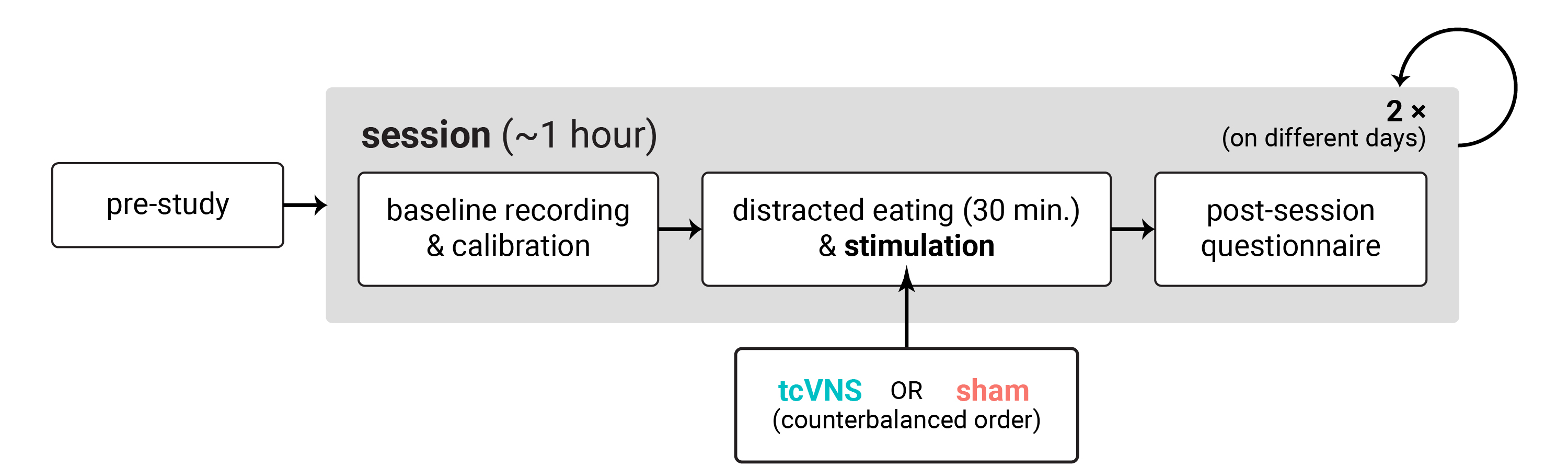}
    \caption{\textbf{Overview of study procedure.}
    After a pre-study phase in which participants select their snack (kept the same for both sessions), participants completed two counterbalanced sessions on different days (approximately 1 hour each). Each session included baseline recording and calibration, a 30-minute distracted eating phase with either tcVNS or sham stimulation, and a post-session questionnaire.}
    \Description{A flow diagram shows a pre-study phase leading into two study sessions on different days. Each session includes baseline recording and calibration, a distracted eating and stimulation phase, and a post-session questionnaire. The stimulation phase is labeled as either tcVNS or sham in counterbalanced order.}
    \label{fig:simplified-procedure}
\end{figure}

The study was conducted over two sessions, each lasting one hour, on separate days. Each participant received both VNS and sham stimulation, with order counterbalanced across sessions to minimize order effects. The study was conducted in a quiet study room with only the participant and the experimenter present. Participants used their own laptops to watch videos and while eating snacks provided by the experimenter. Figure~\ref{fig:simplified-procedure} summarizes the study procedure.

\subsubsection{Snack Selection and Random Assignment}

To balance participants' preferences with experimental control, participants selected one of four snacks before their first study session: chips (Lay's), candy (M\&Ms), cookies (Oreos), or fruit (raisins). These categories represent the most frequently consumed snack types in the U.S. according to the 2024 IFIC survey \cite{insight_2024_2024}. Within each category, we selected the leading brand based on market consumption; however, raisins were used as a shelf-stable substitute for fresh fruit to ensure consistent quality and freshness across all sessions. Participants consumed the same snack across both study sessions. Chips and fruit were each selected by eight participants, while candy and cookies were chosen by four participants each.

Snack portions were standardized to ensure ecological validity, representing an upper-bound consumption level. Following surveys on snacking behavior by Rippin et al. \cite{rippin_comparison_2019}, portions were set at $\text{Mean} + 2\text{SD}$ of a typical snacking episode. Theoretically, this allocation should satisfy the appetite of 97.6\% of participants. These standardized amounts were 59g (chips), 79g (candy), 97g (cookies), and 70g (raisins), all served on a single plate. However, actual consumption frequently reached these limits, which we discuss in Section \ref{limitations}.

% All participants received a standardized amount of the snack they chose. To accurately represent the amount of snacks that would be eaten in a snacking session outside the lab, participants were offered an amount in grams based on the $Mean  + 2 Std.Dev.$ consumed in a single snacking episode based on past surveys by Rippin et al. \cite{rippin_comparison_2019}. In theory, this would enable 97.6\% of participants to have an accurate amount of snacks. These amounts corresponded to 59 grams for chips, 79 grams for candy, 97 grams for cookies, and 70 grams for raisins. All participants received the snacks on a single plate. In practice, more participants than expected finished all the snacks. We discuss potential reasons further in Section \ref{limitations}.

Participants were randomly assigned to one of two groups, counter-balanced by the snack type. In the first group, participants received sham stimulation in the first session and VNS in the second session. Meanwhile, participants in the second group received VNS in the first session, followed by sham stimulation in the second session. Across sessions and groups, participants received identical instructions that they would be receiving electrical stimulation on their neck or shoulder, but not whether it was VNS or sham stimulation. They were also informed that the study concerned eating behavior, but not how eating behavior was expected to be affected.

Participants were asked not to eat anything and not to drink coffee for at least 3 hours prior to the study session. This is common in eating studies, and was designed to ensure all participants would have sufficient appetite. Coffee was added to the fasting requirement to prevent a lingering effect on heart rhythm, which could affect the physiological outcomes.

\subsubsection{Setup and Calibration}

During their first session, participants were asked to provide written informed consent and complete a demographics questionnaire. For both sessions, participants donned a Polar H10 heart rate monitor before completing the pre-study questionnaire. To ensure sufficient time for a heart rate and HRV baseline, this section lasted at least 5 minutes.

Next, the experimenter calibrated the strength of tcVNS or sham stimulation by iteratively increasing the stimulus intensity, stopping at the lowest level at which participants reported a sensation. The intervention was administered in a single-blind manner. During each session, participants were told they would receive neck or shoulder stimulation, but not whether it was tcVNS or sham stimulation. As described in Section \ref{approach}, gel was placed on the stimulator electrodes prior to stimulation, and the stimulator was held in place by the experimenter.

\subsubsection{Distracted Eating \& Stimulation}

For the main section of the study, participants were asked to eat snacks while watching a video of their preference as a distraction. Participants chose a diverse range of video platforms, including YouTube, Netflix, Viki, Bilibili, Hulu, and Prime Video during the study based on their personal interests. This structure was intended to simulate a natural snacking pace, ensure participants were interested in the video, and reduce attention to both the food and the stimulation. Participants were offered a plate of their chosen snack and one cup of water. Participants were instructed that they could eat as much of the snacks on the plate as they wished during the next 30 minutes, but did not need to finish the snacks. 

During this section, participants received tcVNS or sham stimulation on their neck or shoulder respectively. The first period of stimulation started concurrently with the start of the eating section. Each stimulation period lasted two minutes, followed by two minutes without stimulation, for up to a total of five periods. Participants who finished eating earlier than 20 minutes received fewer than five periods. Overall, most participants received the full 10-minute stimulation duration for both tcVNS ($M=9.42, SD=1.50$) and sham ($M=9, SD=1.62$) stimulation. Participants consistently received stimulation at the same strength level, as determined by the calibration in the previous stage for both tcVNS ($4.38, SD=1.01$) and sham stimulation ($M=4.65, SD=0.79$).

Participants were not asked to attend or respond to the stimuli during eating. The sham condition was designed to control for device contact, handling, and tactile sensation while applying stimulation away from the cervical vagus nerve.

This section of the study ended either after 30 minutes had elapsed (23 sessions), after participants had eaten all the snacks on their plate (21 sessions), or when participants indicated they were full early (4 sessions).

\subsubsection{Post-Experiment}
Finally, after the participants finished eating, they completed the post-study questionnaire, including qualitative feedback about the experience. They took off the Polar H10 monitor and were offered dry wipes to remove gel. Between each study, the stimulator was sanitized with wet wipes and fully recharged.

\subsection{Statistical Analyses}
Hierarchical linear mixed-effects models (LMEs) were used to conduct statistical analyses for the quantitative measures collected in the study, using the \textit{lme4} package \cite{lme4} in R \cite{R}. LME models can handle unbalanced classes and missing data, providing a more robust analysis while being equivalent to a repeated-measures ANOVA in simple cases. For models with multiple comparisons (e.g. interactions) that showed significant effects, we conducted post-hoc tests to estimate mean differences ($\Delta M$) using the \textit{emmeans} package, using Bonferroni corrections.

The study was powered for within-subject comparisons between tcVNS and sham stimulation. Interaction terms, including condition-by-snack-type effects, were treated as exploratory because the study was not powered to estimate subgroup-specific effects. All models included participant as a random intercept, consistent with the repeated-measures design and power analysis. For exploratory models with additional factors, we fit nested condition-only, additive (condition + other main effect), and interaction models (condition * other main effect), inspected fixed-effect terms, and compared model fit using AIC. Video type was coded as either informational (e.g. lectures, technology, finance) or entertainment (e.g. TV shows, comedy) and examined in exploratory models. However, video platform was too sparse for analysis, so we did not include it.

\section{Results}

In this within-subject study, participants ate less and took longer to eat when receiving vagus nerve stimulation (VNS) rather than sham stimulation. Overall, participants felt satiated to the same degree in both conditions despite eating less with VNS. Participants felt a high sense of control but low embodiment with the stimulator regardless of condition. Moreover, participants had a lower heart rate during VNS, suggesting autonomic activity, but there was no significant difference in heart rate variability (HRV). Finally, participants felt the stimulation was subtle, like a mild vibration.

\subsection{Behavioral Measures}

\begin{figure}
    \centering
    \includegraphics[width=\linewidth]{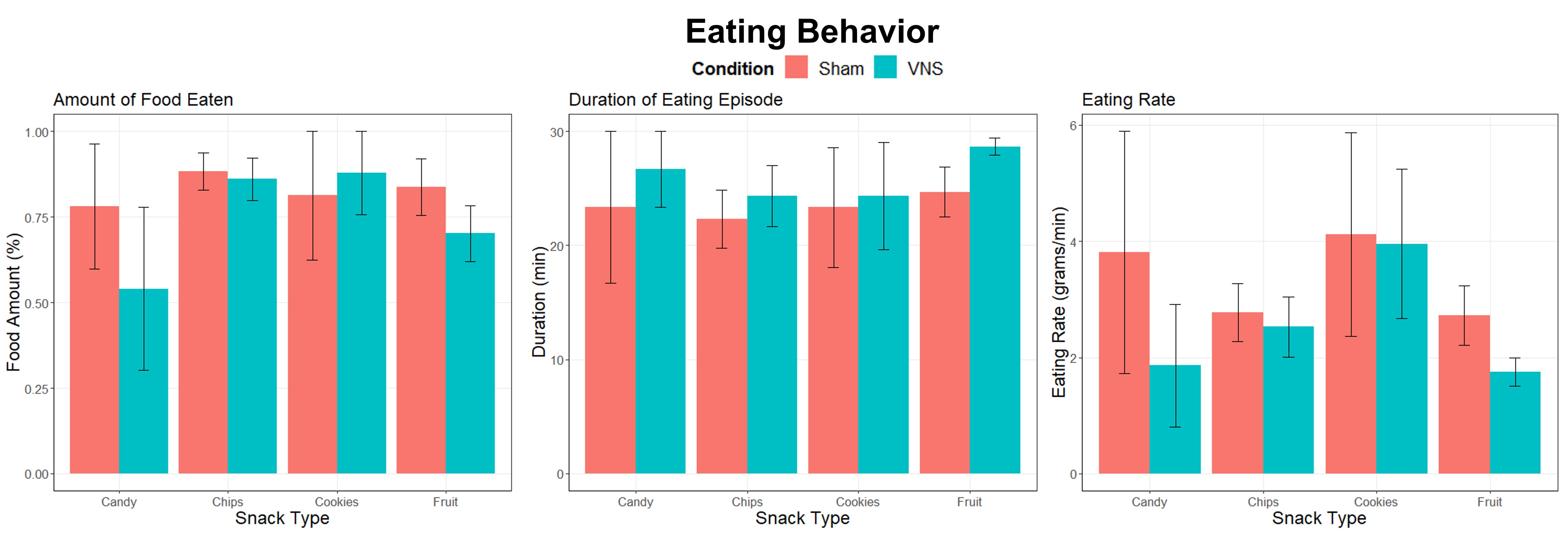}
    \caption{\textbf{Effects of stimulation on eating behavior across snack types.} Mean amount of food eaten (left), duration of the eating episode (center), and eating rate (right) are shown for sham and tcVNS conditions across four snack types. Across snacks, participants ate less food, took longer to finish eating, and ate more slowly under tcVNS compared to sham stimulation.}
    \Description{Three grouped bar charts compare sham and VNS across candy, chips, cookies, and fruit. The charts show amount of food eaten, duration of the eating episode, and eating rate. Across snack types, VNS generally corresponds to lower intake, longer duration, and slower eating rate than sham.}
    \label{fig:eatingbehavior}
\end{figure}

For each behavioral measure, we tested main effects of experimental condition and snack type, as well as their interaction, by constructing a set of LMEs. We increased model complexity iteratively, first evaluating main effects independently before investigating interactions. Only the effect of condition was considered as a primary outcome, whereas the effect of snack type was exploratory, due to the limited number of data points per snack type. The changes due to both condition and snack type are shown in Figure \ref{fig:eatingbehavior}.

% Amount
\textbf{Food intake was significantly lower under tcVNS.}
We found that only the condition (\(\beta = -0.08\), \(95\%\ CI = [-0.15, -0.02]\), \(t = -2.52\), \(p = 0.016\)) had a significant main effect on the amount of food eaten in the study. Neither chips (\(p=0.174\)), cookies (\(p=0.329\)), nor fruit (\(p=0.481\)) had a significant main effect on food eaten. The interaction model had lower AIC than the condition-only and additive models (AIC=-18.83 vs. -18.00 and -14.59), with significant interaction terms for chips (\(\beta = 0.22, 95\%\ CI = [0.03, 0.40], t = 2.40, p = 0.026\)) and cookies (\(\beta = 0.31, 95\%\ CI = [0.08, 0.53], t = 2.75, p = 0.012\)). The results indicated that participants ate significantly less under tcVNS than under sham stimulation.

Post-hoc comparisons confirmed the primary finding, indicating that participants ate significantly less under tcVNS than under sham stimulation (\(\Delta M = 0.081, t = 2.52, p = 0.019\)). However, post-hoc pairwise comparisons showed that the interaction between condition and snack type was not significant (\(ps > 0.999\), except VNS:Candy-Chips, where \(p=0.34\) and VNS:Candy-Cookies, where \(p=0.58\)). These findings indicate that food intake was lower overall under tcVNS, but do not provide reliable evidence that the effect differed by snack type. The estimated mean differences correspond to a 9.6\% reduction in food intake.

% Duration
\textbf{Eating episodes were significantly longer under tcVNS.}
For the duration of the eating episode, we found that only the condition (\(\beta = 2.79, 95\%\ CI = [0.30, 5.29], t = 2.26, p = 0.029\)) had a significant main effect. None of the snack types (\(p=0.696\), \(p=0.823\), \(p=0.696\)) had a main effect on eating duration, and there were no significant interactions between condition and snack type (\(p=0.756\), \(p=0.657\), \(p=0.876\)). Model comparison favored the condition-only model over additive and interaction models including snack type (AIC=315.35 vs. 319.79 and 324.94). Overall, eating episodes were significantly longer under tcVNS by an estimated 2.79 minutes, or by 11.5\% of the original duration.

% Eating Rate
\textbf{Eating rate was significantly slower under tcVNS.}
Similar to duration, only condition (\(\beta = -0.72, 95\%\ CI = [-1.30, -0.14], t = -2.484, p = 0.017\)) had a main effect on eating rate. None of the snack types (\(p=0.86\), \(p=0.37\), \(p=0.58\)) had an effect on eating rate, and there were no significant interactions between condition and snack type (\(p=0.072\), \(p=0.121\), \(p=0.293\)). Model comparison favored the condition-only model over additive and interaction models including snack type (AIC=181.29 vs. 184.07 and 185.48). Overall, these results indicate that participants ate significantly more slowly under tcVNS than under sham stimulation. The effect was an estimated 0.719 grams per minute, or a 23.6\% reduction in eating speed.

\textbf{Video type did not explain behavioral outcomes.}
We also explored whether participants' self-selected video type (informational vs. entertainment) affected the behavioral outcomes. For food intake, video type did not have a significant main effect (\(p=0.542\)), nor did it interact with condition (\(p=0.287\)); the condition-only model also had lower AIC than additive and interaction models including video type (AIC=-18.00 vs. -16.01 and -15.31). Similarly, video type had no significant main effect on eating duration (\(p=0.272\)) or eating rate (\(p=0.266\)), and neither outcome had a significant condition \(\times\) video type interaction (\(p=0.111\) and \(p=0.321\), respectively).

\subsection{Self-Reported Measures}

\begin{figure}
    \centering
    \includegraphics[width=\linewidth]{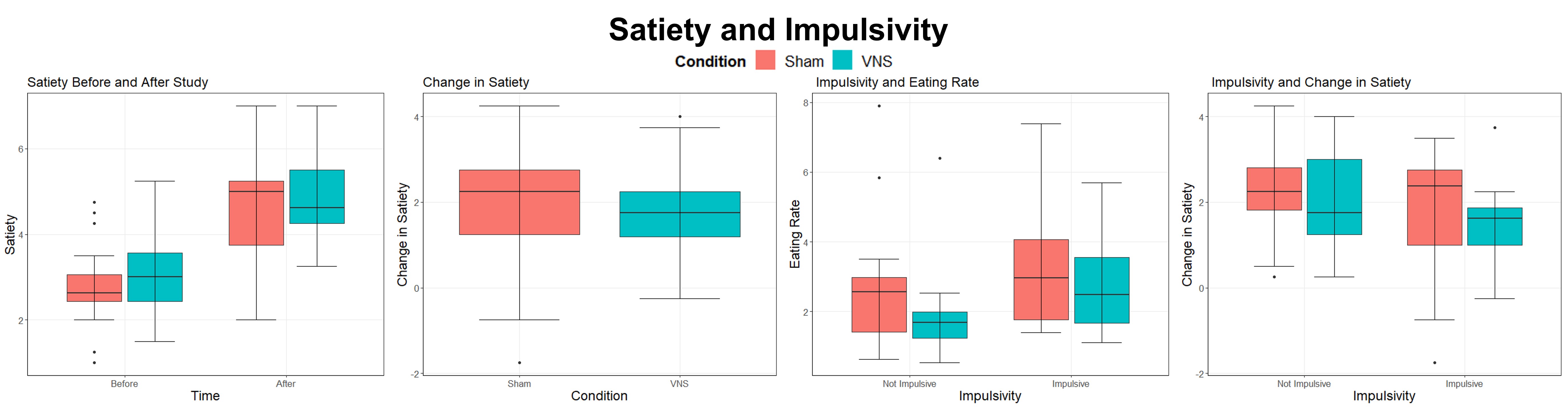}
    \caption{\textbf{Satiety and impulsivity under sham and tcVNS conditions.} Self-reported satiety increased significantly after the eating episode in both conditions, with no difference between sham and tcVNS, indicating that reduced intake under tcVNS did not diminish subjective satiety. Eating impulsivity did not significantly affect eating rate or change in satiety, nor did it interact with stimulation condition.}
    \Description{Four box plots summarize satiety and impulsivity results. Satiety increases from before to after eating in both sham and VNS conditions. Change in satiety is similar across conditions. Eating impulsivity does not show a clear effect on eating rate or change in satiety.}
    \label{fig:satiety-impulsivity}
\end{figure}

For self-reported quantitative measures, we tested the main effects of condition and the eating episode, as well as their interaction. We also explored whether impulsivity was associated with our primary measures, eating rate and satiety, given the relevance of impulsive eating behavior for future clinical or at-risk populations. Lastly, we explored how the experimental condition affected participants' perceptions of the stimuli, as well as their agency and embodiment within the system. Once again, we increased model complexity iteratively, first evaluating main effects independently before investigating interactions. The changes in satiety during the experiment, as well as the relationship of impulsivity with other variables, are shown in Figure \ref{fig:satiety-impulsivity}.

% Satiety before and after
\textbf{Satiety was higher after snacking in both conditions.} 
The experimental condition did not have a significant effect on satiety (\(p=0.443\)). Meanwhile, the timing of the satiety questionnaire, i.e., whether it was before or after the eating episode, had a significant effect on satiety (\(\beta = 1.86, 95\%\ CI = [1.48, 2.25], t = 9.58, p < 0.001\)). There was no significant interaction between the condition and timing (\(p=0.633\)), implying that participants felt satiated to the same degree after snacking in both the sham and VNS conditions.

% Eating impulsivity
\textbf{Eating impulsivity had no significant effect on eating rate or change in satiety.}
As described in Section \ref{measures}, we treated impulsivity as a binary factor for statistical analysis. Although we also attempted to investigate relationships directly with the impulsivity scale, there was no discernible pattern or correlation with any primary outcome. Impulsivity did not significantly affect eating rate (\(p = 0.747\)) and had no interaction with experimental condition (\(p = 0.204\)). Similarly, impulsivity did not significantly affect change in satiety (\(p = 0.093\)), and did not have any interaction with condition (\(p = 0.849\)) either. These results do not provide evidence that the behavioral effect differed by impulsivity in this healthy sample. However, given the limited participant count and focus on healthy participants, further studies are needed to determine the effects on clinically impulsive populations.

\begin{figure}
    \centering
    \includegraphics[width=0.7\linewidth]{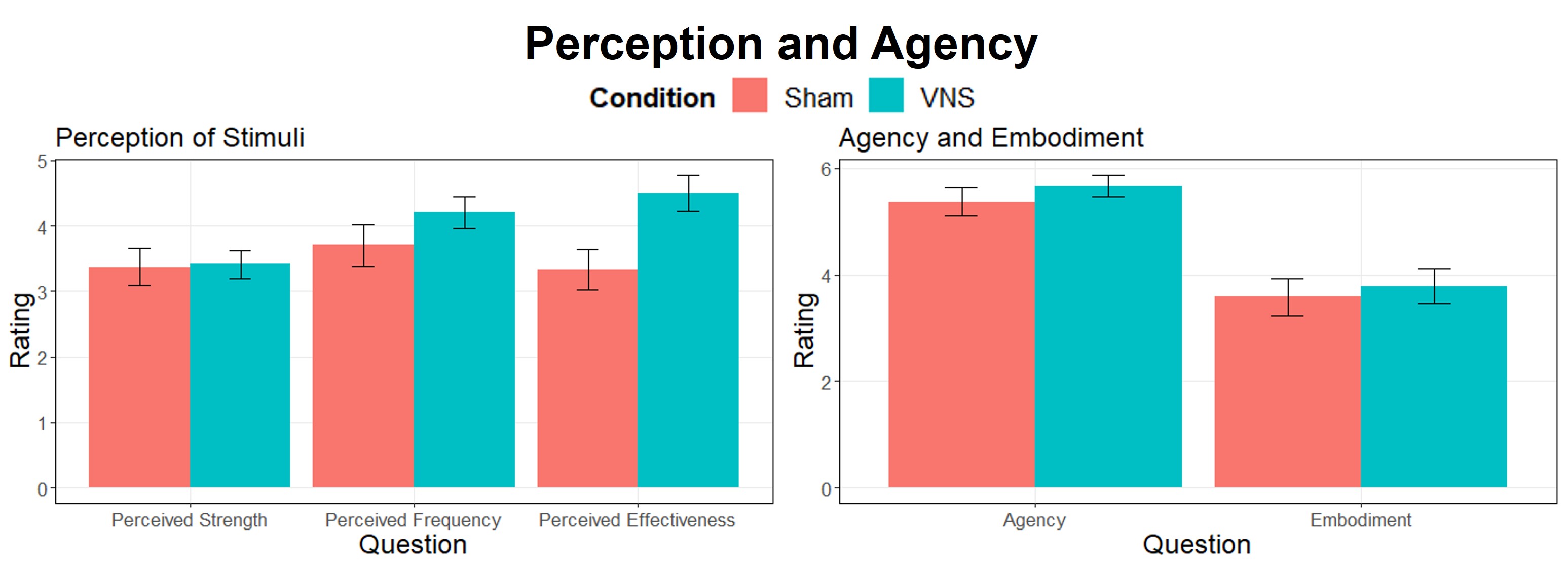}
    \caption{\textbf{Perception of stimulation (left) as well as agency and embodiment (right).} Participants rated tcVNS as affecting their eating behavior more than sham stimulation, despite reporting no differences in perceived strength or frequency of stimulation. Stimulation condition did not significantly affect participants' sense of agency or embodiment, with agency remaining high and embodiment relatively low across conditions.}
    \Description{Two bar charts compare sham and VNS ratings. The first chart shows perceived strength, perceived frequency, and perceived effectiveness of stimulation. VNS has a higher perceived effectiveness rating, while perceived strength and frequency are similar. The second chart shows agency and embodiment ratings, which are similar across conditions.}
    \label{fig:perception-agency}
\end{figure}

% Physical sensation
\textbf{Participants rated tcVNS as affecting their eating more than sham stimulation.} 
The experimental condition had no significant effect on participants' perception of how strong the stimuli were (\(p = 0.89\)). Similarly, there was no significant difference in how often participants perceived the stimuli (\(p = 0.168\)). The perceived strength of stimuli was weak (\(M = 3.39\), \(SD = 1.21\)) while the perception frequency was average (\(M = 3.96\), \(SD = 1.38\)), confirming that the sensation was mild. However, participants rated tcVNS as affecting their eating more than sham stimulation (\(\beta = 1.17, 95\%\ CI = [0.55, 1.78], t = 3.83, p < 0.001\)). This was a surprising result, considering the study protocol was nearly identical across conditions, with the only change being the site of stimulation. It's important to note that this question was not directional, and some participants later remarked they thought tcVNS was supposed to increase their appetite rather than decrease it. In all, participants perceived a difference in how the two conditions affected their eating, but the physical sensation of stimulation did not otherwise differ from sham stimulation.

Participants' sense of agency (\(p = 0.435\)) over their eating behavior, and their embodiment with the device (\(p=0.356\)) were not significantly affected by the experimental condition. Yet, the overall sense of agency was high in both conditions (\(M = 5.52\), \(SD = 1.15\)), implying that the participants felt high control over their eating. The sense of embodiment with the device was low (\(M = 3.69\), \(SD = 1.65\)). Some participants remarked that this was due to the device being handheld and held by the experimenter, whereas it would feel more natural if it were wearable. The perception of the stimuli and participants' sense of agency and embodiment with the system are shown in Figure \ref{fig:perception-agency}.

\subsection{Physiological Measures}

\begin{figure}
    \centering
    \includegraphics[width=\linewidth]{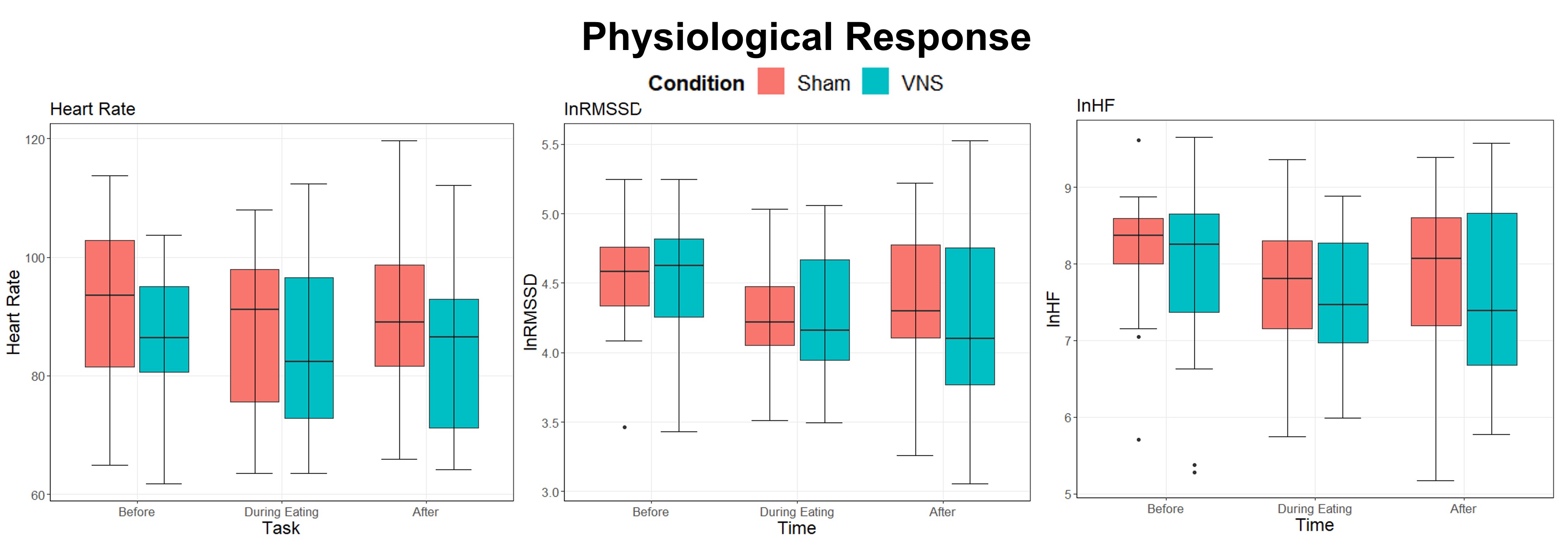}
    \caption{\textbf{Physiological responses to eating under sham and tcVNS conditions.} 
    Heart rate (left), lnRMSSD (center) and lnHF (right) are shown across three time periods: before eating, during eating, and after eating, for sham and tcVNS conditions. Participants exhibited lower heart rate during tcVNS, while both lnRMSSD and lnHF decreased during eating regardless of condition.}
    \Description{Three box plots compare physiological responses for sham and VNS before eating, during eating, and after eating. The plots show heart rate, lnRMSSD, and lnHF. Heart rate is lower under VNS, while lnRMSSD and lnHF do not show a clear condition difference and decrease during eating.}
    \label{fig:phys-response}
\end{figure}

For the physiological measures, we tested the main effects of condition and time (divided into before eating, while eating, and after eating), as well as their interaction. For the HRV metrics, we used their natural logarithm, lnRMSSD and lnHF as specified in Section \ref{measures}. Yet again, we increased model complexity iteratively, first evaluating main effects independently before investigating interactions. The changes in heart rate and HRV during the experiment and by experimental condition are shown in Figure \ref{fig:phys-response}.

% HR
\textbf{Participants had significantly lower heart rate under tcVNS.}
The experimental condition had a significant effect on participants' average heart rate (\(\beta = -5.16, 95\%\ CI = [-8.74, -1.58], t = -2.85, p = 0.005\)), which was lower under tcVNS. The eating task did not have a significant effect on heart rate (\(p = 0.339\) while eating; \(p = 0.453\) after eating), nor was there an interaction between condition and the eating task (\(p = 0.913\) while eating; \(p = 0.678\) after eating). A decrease in heart rate can be consistent with increased parasympathetic nervous activity~\cite{shafferOverviewHeartRate2017}, aligning with the predicted direction of tcVNS. Food intake is typically associated with a short-term increase in heart rate, but this was not observed in our study, perhaps because the snack amount was too small or the effect was too small.

% HRV - RMSSD
\textbf{RMSSD was significantly lower while eating, but was not affected by tcVNS.}
For the root mean square of successive differences (RMSSD) metric, the most commonly used HRV measure, the experimental condition had no significant effect (p = 0.56). On the other hand, participants had significantly lower RMSSD while they were eating (\(\beta = -0.23, 95\%\ CI = [-0.42, -0.04], t = -2.44, p = 0.017\)) and after eating (\(\beta = -0.19, 95\%\ CI = [-0.38, 0], t = -1.98, p = 0.05\)). There were no significant interactions between the eating task and condition, during eating (\(p = 0.920\)) or after eating (\(p = 0.938\)).

Post-hoc pairwise tests showed that RMSSD was significantly lower only during eating (\(\Delta M = 0.23, t = 2.44, p = 0.043\)), not after eating (\(\Delta M = 0.19, t = 1.98, p = 0.12\)). RMSSD can decrease temporarily during eating due to postprandial physiological changes, so this result was expected. Although tcVNS is often associated with increased RMSSD, we did not observe any effect on RMSSD in the study. There are many potential causes for this effect, including that the number of participants was too small, the effect was obscured by the physiological response to eating, or due to the experimental setting, as we further discuss in Section \ref{limitations}. 

% HRV - HF
\textbf{HF was not significantly affected by tcVNS or the eating task.}
High-frequency (HF) heart rate variability was also not significantly affected by the experimental condition (\(p = 0.267\)). Similar to the results for RMSSD, HF was significantly lower while eating (\(\beta = -0.4, 95\%\ CI = [-0.77, -0.02], t = -2.1, p = 0.038\)), but not after eating (\(p = 0.095\)). There were no significant interactions between the eating task and condition, during eating (\(p = 0.77\)) or after eating (\(p = 0.82\)).

Post-hoc pairwise tests showed that HF was not significantly different during eating (\(\Delta M = 0.40, t = 2.10, p = 0.095\)), or after eating (\(\Delta M = 0.31, t = 1.69, p = 0.21\)). HF is often interpreted in relation to vagal activity, but past studies have also shown HF to be reduced after eating~\cite{sauderEffectMealContent2012}. As HF is a frequency domain HRV feature, it requires longer-term data collection to be reliable. In addition to the issues mentioned for RMSSD, the time period in our study may have been insufficient to sufficiently reduce noise in HF measurements.

Altogether, the physiological measures provide mixed evidence of autonomic response. Heart rate was lower during tcVNS, while HRV measures did not differ by condition, and RMSSD decreased during eating. These results suggest physiological engagement during the study, but do not establish increased vagal tone or a specific autonomic mechanism for the observed behavioral effects.

\subsection{Qualitative Analysis}

%    * Categories for whether they would use or not (which is quite interesting: a lot of participants do not perceive impact or did not infer use).
%           > Maybe cross check if it produced an effect.
%           > Baseline healthiness? Eating impulsivity (average of three impulsivity questions).

\begin{figure*}[t]
    \centering
    \includegraphics[width=\textwidth]{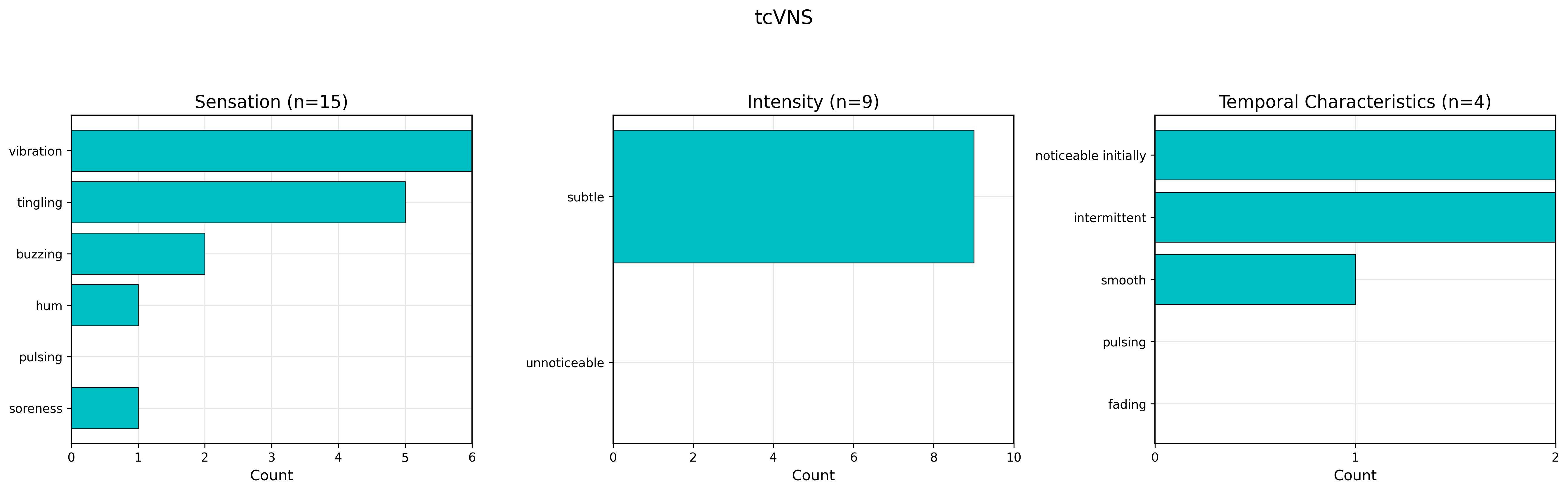}
    \vspace{0.75em}
    \includegraphics[width=\textwidth]{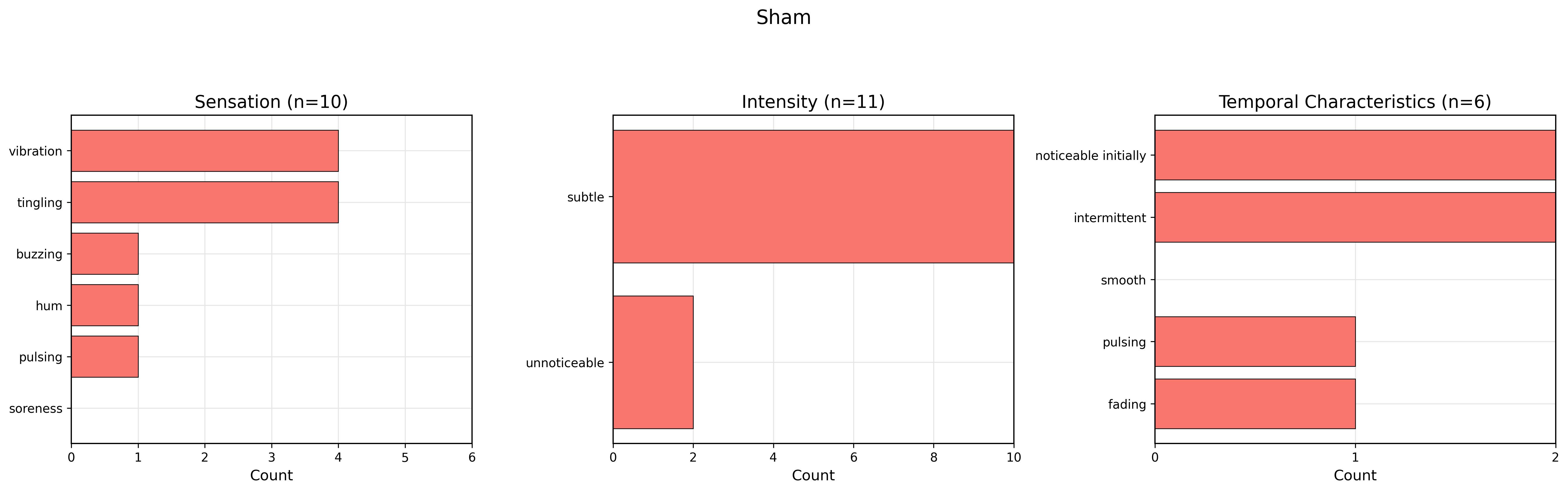}
    \caption{
    \textbf{Qualitative sensory descriptors under sham and tcVNS condition.} 
    Distributions of participant-reported descriptors for \textit{sensation}, \textit{intensity}, and \textit{temporal} are shown for tcVNS (top) and sham (bottom) conditions. Counts indicate number of participants who mentioned each descriptor (or a similar term) at least once. The number of participants contributing to each subplot ($n$) is reported as not all participants provided qualitative feedback covering all dimensions.
    }
    \Description{Two rows of horizontal bar charts show participant-reported sensory descriptors for tcVNS and sham. For tcVNS, vibration and tingling are the most common sensation descriptors, subtle is the dominant intensity descriptor, and noticeable initially and intermittent are the most common temporal descriptors. For sham, vibration and tingling are also common, subtle is the dominant intensity descriptor with some unnoticeable reports, and noticeable initially and intermittent are the most common temporal descriptors.}
    \label{fig:qualitative_descriptors}
\end{figure*}

\subsubsection{Sensory Analysis}
Figure~\ref{fig:qualitative_descriptors} summarizes participant-reported sensations for tcVNS and sham conditions across sensation, intensity, and temporal dimensions.

\textbf{Sensation.} In both conditions, participants most commonly described stimulation as vibrating or tingling. tcVNS elicited a wider range of secondary sensations (e.g., buzzing, hum, soreness), whereas sham produced fewer descriptors.

\textbf{Intensity.} Perceived intensity was generally low. tcVNS was most often perceptible but subtle, while sham more often approached being unnoticeable.

\textbf{Temporal characteristics.} Temporal descriptions were similar across conditions, with sensations often noticeable only at first and fading over the session.

\textbf{Likeness.} Everyday analogies were infrequent; when provided, participants ($n=4$) most often likened the sensation to a massage gun (P8, P17, P21) or car vibration (P7).

\textbf{Secondary impacts.} More participants contributed impact-related comments under tcVNS ($n=11$) than sham ($n=6$). tcVNS impacts included relaxation or calmness (P23, P21, P14), slight nausea (P1), discomfort (P11, P12, P18), and dry mouth (P13), whereas sham impacts were fewer and more uniformly negative. Appetite-related comments were rare ($n=2$ per condition) and mixed.

\textbf{Overall.} Both conditions were primarily low-intensity and vibration-like, but tcVNS elicited a broader experiential profile.

\subsubsection{Form-Factor Feedback}
A subset of participants in both the tcVNS ($n=6$) and sham ($n=9$) conditions commented on the device form factor or manual application by the experimenter. Several participants reported pressure-related discomfort, including interference with swallowing (P2) or movement while eating (P3), and some expressed unease with having the device held by another person (P21, P25). Similar concerns appeared in both conditions, suggesting that constraints on movement and posture arose from manual application rather than stimulation at a specific site. Across conditions, participants preferred a more autonomous or wearable form factor, such as a wearable attachment or electrodes (P10, P24), highlighting the importance of ergonomics for future tcVNS studies.

\subsubsection{Adoption in Daily Life} 
We focused our analysis on the tcVNS condition. Twenty-one participants expressed a clear adoption stance. Of these, two indicated they would use the device in daily life, eleven expressed conditional willingness to adopt, and eight indicated they would not use the device.

\textbf{Conditional adoption.} Conditional willingness to adopt tcVNS was the most common stance. The dominant factor shaping adoption was perceived efficacy: seven participants said they would consider using the device only if it demonstrably affected their eating behavior or appetite, such as ``if it helps me eat healthier'' (P7) or ``if there is actual proven benefits'' (P11). Form factor also mattered, with two participants preferring a wearable or non-hand-held design and one emphasizing that the device would need to look suitable for everyday wear, such as ``smart jewelry'' (P7). One participant cited lack of perceived benefit as a reason for hesitation (P19), and two expressed conditional willingness without specifying clear reasons.

\textbf{Non-adoption.} Eight participants indicated they would not use the device in daily life. Reasons included negative impact on the eating or viewing experience (P2, P6), aversion to the sensation itself (P8), lack of perceived benefit (P9, P26), and values-based resistance to controlling hunger with an external device (P13). These responses suggest that non-adoption was primarily driven by absence of perceived utility rather than strong negative reactions.

\textbf{Intended use.} Seven participants articulated how they might use tcVNS if adopted. Most described use cases related to eating regulation ($n=6$), including avoiding further eating after reaching satiety, reducing snacking between meals, or supporting healthier eating habits. One participant described using the device to support focus during tasks (P11).

\textbf{Uncertainty about purpose.} Notably, three participants expressed uncertainty about the purpose or benefits of the device, independent of their adoption stance. These participants explicitly reported not understanding what the stimulation was intended to do: ``I am not sure what [the device is] used for or what are its benefits'' (P1); ``I don't know why it's used'' (P24).

\section{Discussion}

Our study evaluated AppetiteCheck, a novel eating intervention using momentary tcVNS to influence eating behavior. We assessed the effect of momentary tcVNS on eating behavior, satiety, user experience, and autonomic nervous system activity. We combined quantitative analyses of behavioral data, questionnaires, and sensors with open-ended questions to gather qualitative feedback and build a comprehensive picture of AppetiteCheck. In this section, we discuss the feasibility of AppetiteCheck relative to prior work, explore the design space of interventions targeting autonomic processes, and consider the potential of tcVNS as a physiological intervention.
% Sensory analysis of qualitative feedback seems to confirm that the vagus nerve stimulation is subtle (not strong or overly distracting, etc.).
% Analysis of \textit{Would you use this device in your daily life? If so, how?} suggests that participants often did not consciously perceive the impact of the stimulation, even if it did impact them, which is exciting.

\subsection{Feasibility and Effectiveness of Momentary tcVNS}

%    \item[\textbf{RQ1:}] How does momentary tcVNS affect the amount of food eaten and eating rate in a single eating episode?
    % \item[\textbf{RQ2:}] How is satiety after an eating episode affected by momentary tcVNS?
    % \item[\textbf{RQ3:}] What is the effect of momentary tcVNS on the autonomic nervous system?
% Rapid appetite change, no change in satiety, parasymathetic activation

Participants ate fewer snacks over a longer period during tcVNS than during sham stimulation, resulting in a slower eating rate. Thus, answering our RQ1, momentary tcVNS was associated with less eating and slower eating during a single controlled snacking episode. Quantifying this effect, we found that participants ate 9.6\% less and ate 23.6\% slower. In this task, these differences were numerically larger than device-based mindful eating approaches like SSpoon, where participants ate 4.5\% less and 14.9\% slower~\cite{chen_sspoon_2022}. The intake reduction was also numerically similar to that of low doses (0.75pmol/kg/min) of intravenous GLP-1 infusions, which reduce food intake by 9.5\%, but weaker than larger doses (1.5 pmol/kg/min), which reduce food intake by 34.2\%~\cite{gutzwillerGlucagonlikePeptide1Potent1999}. Therefore, AppetiteCheck shows promise as a device-based intervention for reducing intake and slowing eating during a single eating episode.

For RQ2, satiety did not differ significantly between sham stimulation and tcVNS, but increased after an eating episode regardless. Unlike studies on semaglutide, which show increased satiety \cite{gibbonsEffectsOralSemaglutide2021}, AppetiteCheck did not increase post-eating satiety in this short-term study. However, participants reported comparable satiety while eating less, suggesting that reduced intake did not come at the cost of lower immediate post-eating satiety. Because we assessed satiety immediately after a single snacking episode, these findings do not establish how tcVNS would affect appetite or satiety across a full day or over repeated use.

Lastly, for RQ3, tcVNS was associated with reduced heart rate, consistent with physiological engagement but not diagnostic of parasympathetic mediation. The physiological results were mixed, as HRV did not differ significantly between tcVNS and sham stimulation for either RMSSD or HF. We therefore cannot determine from these data whether the behavioral effects were mediated by increased parasympathetic activity. A longer deployment study with more continuous physiological measurement would be needed to characterize the autonomic pathway more clearly.

Altogether, our findings show that momentary tcVNS reduced food intake and eating rate relative to sham stimulation in a controlled snacking task with healthy participants. The observed effect was among the larger non-pharmacological effects in our comparison, while post-eating satiety was maintained despite lower food intake. At the same time, the study does not establish longer-term effectiveness, effects during regular meals, or the physiological pathway linking stimulation to behavior change.

For ubiquitous computing, these findings suggest a new direction for implicit health interventions: using momentary physiological intervention rather than perceptual cues that users must notice, interpret, or follow. For eating systems more broadly, AppetiteCheck also shows how eating episode detection could be paired with a low-effort intervention delivered during eating, rather than ending at sensing or self-tracking. However, this shift creates a design challenge: when an intervention is subtle, users may not know whether it is active, effective, or appropriate for the moment. Future closed-loop physiological interventions may therefore need interaction designs that preserve user control and make activation accountable, even when the intervention itself remains subtle and undemanding.

\subsection{Towards Closed-Loop, Wearable tcVNS}
\label{wearable-discussion}

Our study explored momentary tcVNS in controlled lab conditions; real-world use would require further evaluation of safety, stimulation paradigm, and effectiveness. Based on our qualitative findings, a promising path for AppetiteCheck is a wearable rather than a handheld device, which better aligns with its low-effort, peripheral design. We therefore examine power, wearable form factors, and closed-loop deployment.

\subsubsection{Power and Operational Lifetime}

We profiled the Truvaga stimulator at the highest stimulation level used in our study with an oscilloscope and a 500$\Omega$ IEC test load~\cite{iec60601_2_10_2023}. This produced a peak voltage of 10.64V, RMS voltage of 7.52V, and peak current of 21.3mA. The sinusoidal burst pattern has a 2.5\% duty cycle (1ms bursts delivered at 25Hz). Assuming 75\% DC-DC boost efficiency, this yields an estimated average battery draw of $((7.52)^2/500\Omega) \times 2.5\% / 75\% = 3.77mW$ during stimulation, or $0.126mWh$ per 2-minute stimulation period and $0.628mWh$ for the 10-minute period used in our in-lab study.

% This is higher than typical non-invasive nerve stimulation levels (~5mA) as a result of the sinusoidal burst approach Truvaga uses, in contrast with standard square waves, which typically makes the stimulation more tolerable \cite{daliComparisonEfficiencyChopped2019}.

We estimate operational lifetime for three deployment approaches: on-demand tcVNS with only Bluetooth communication, tcVNS with IMU and PPG for low-power eating detection~\cite{thomazPracticalApproachRecognizing2015}, and tcVNS + MunchSonic for high-fidelity eating detection~\cite{mahmudMunchSonicTrackingFinegrained2024}. In all cases, we assume 24 sessions (48 minutes) per day, corresponding to $3.77mW \times 48min  / 60 min = 3.02 mWh$. By default, we assume a 110mAh Li-ion battery (approximately 407mWh at 3.7V) based on prior neck-based wearables in a jewelry form factor~\cite{xueECGNecklaceLowpower2025,biAuracleDetectingEating2018,amoresDevelopmentStudyEzzence2022}; for the tcVNS + MunchSonic variant, we also consider a 350mAh battery based on BioEssence~\cite{amoresBioEssenceWearableOlfactory2018}. Daily power consumption is calculated as $P_{sense}\times23.2h + (P_{sense}+P_{stim})\times0.8h$, where $P_{stim}=3.77mW$.

For on-demand tcVNS, the stimulator would not perform sensing. With a low-power BLE microcontroller such as the nRF5340, it would advertise passively during the day and transmit data only during stimulation, mapping to approximately 0.18mW and 30mW based on Nordic documentation and power profiles~\cite{nordicNRF5340ProductBrief2026,nordicOnlinePowerProfiler2026}. This configuration would last 313 hours, or 13 days, suggesting that a compact, jewelry-scale stimulator could be recharged weekly.

For closed-loop tcVNS with low-power sensing, we use a conservative 4mW estimate for IMU+PPG+BLE, informed by PPG Earring's 2.2mW estimate for PPG+accelerometer+BLE~\cite{xuePPGEarringWireless2025}. This configuration would last approximately 99 hours, or 4.1 days -- sufficient for multi-day use.

For tcVNS with MunchSonic-style active acoustic sensing, we use MunchSonic's reported 96.5mW sensing power~\cite{mahmudMunchSonicTrackingFinegrained2024}. Under the same conditions, it would last only 4.2 hours with a 110mAh battery. A larger 350mAh battery (approximately 1,295mWh) would extend this to 13.4 hours, making partial-day deployment feasible but still falling short. Full-day use may therefore require a hybrid approach, using IMU to detect when the user is active before activating acoustic sensing for detailed eating behavior.

Across scenarios, tcVNS itself contributes only a small fraction of the energy budget; the main power constraint is the sensing modality used to detect eating. Thus, these estimates support energy feasibility, while wearable feasibility still depends on electrode contact, ergonomics, and safety validation.

% To evaluate the feasibility of a self-contained wearable, we modeled the energy budget of a hybrid device that integrates the active acoustic sensing of MunchSonic~\cite{mahmudMunchSonicTrackingFinegrained2024} for automated eating detection with the intermittent neurostimulation of AppetiteCheck. For this calculation, we assume the MunchSonic hardware replaces the generic microcontroller overhead—drawing 96.5 mW for continuous monitoring—and utilizes a standard 500mAh (1,850 mWh) battery. In this ``worst-case'' 24-hour profile, tcVNS stimulation draws an additional 26.6 mW from the battery (factoring in 75\% boost converter efficiency) only during active sessions. 

% Under a high-usage scenario of twenty 2-minute stimulation periods per day (1 hour, consistent with gammaCore safety tests), the stimulation itself represents only 3.5\% (81.2 mWh) of the total daily energy demand. However, the continuous sensing overhead requires 2,252.3 mWh per day, resulting in a total daily demand of approximately 2,333.5 mWh. Consequently, a 500 mAh battery would provide roughly 19.0 hours of operational lifetime, sufficient for full day use if the device was removed and charged during sleep. This indicates that while the neurostimulation mechanism is very energy-efficient for everyday use, the operational bottleneck for a closed-loop system lies in the power draw of the continuous sensing modality.

\subsubsection{Device and Electrode Design}

\begin{figure}
    \centering
    \includegraphics[width=\linewidth]{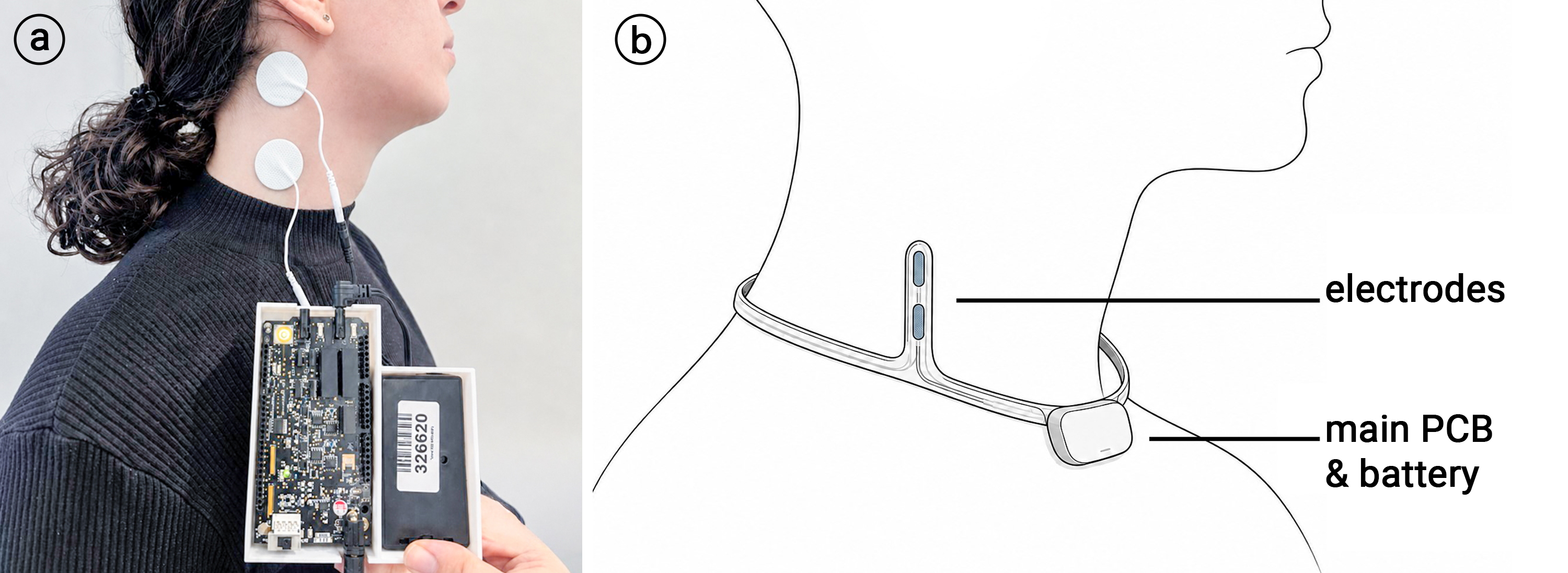}
    \caption{\textbf{Wearable implementation path for AppetiteCheck.} (a) Initial stimulator prototype used to explore wearable tcVNS before in-lab study. (b) Proposed necklace-style form factor that integrates the stimulation module and embedded electrode contacts. The necklace design is a future direction, not a device evaluated in the present study.}
    \Description{Two-panel figure showing a wearable implementation path. The first panel shows an early stimulator prototype with gel electrodes placed on the side of the neck and electronics held near the chest. The second panel sketches a necklace-style device with electrode contacts near the side of the neck and a main PCB and battery module at the front.}
    \label{fig:wearable-design}
\end{figure}

Although our study used an FDA-approved handheld stimulator for safety and physiological validity, the design path does not require a handheld form factor. Our initial AppetiteCheck prototype used the NeuroStimDuino \cite{neuralaxyNeuroStimDuino2026} and was pocket-sized, as shown in Figure \ref{fig:wearable-design}a. For future deployment, our qualitative findings and prior neck-worn sensing systems \cite{xueECGNecklaceLowpower2025,zhangNeckSenseMultisensor2020} suggest a necklace form factor. A pendant-style enclosure could house the battery and electronics near the cervical stimulation site, while a close-fitting choker, shown in Figure \ref{fig:wearable-design}b, or an adjustable wrap or shawl could maintain electrode contact without requiring users to hold or position the stimulator during eating.

Prior work supports the miniaturization of stimulation electronics, with electrical muscle stimulation hardware made small enough to fit inside a smartwatch \cite{takahashiCanSmartwatchMove2024}. AppetiteCheck uses a single stimulation channel and low stimulation power, so the core electronics can be implemented with a small set of components, including a microcontroller, DC/DC converter, high-voltage transistor and diode, supporting resistors and capacitors. This shifts the remaining wearable design challenges toward high-voltage safety, electrode contact, and long-term comfort. Before user deployment, the wearable version would require benchtop verification of current limits, isolation, fault behavior, and stimulation consistency.

A key unresolved problem is electrode design. Recent HCI work on electrical muscle stimulation~\cite{takahashiCanSmartwatchMove2024, knibbeSkillSleevesDesigningElectrode2021} and electrical nerve stimulation~\cite{brooksStereoSmellElectricalTrigeminal2021} has integrated stimulation electrodes into wearables such as smartwatches, sleeves, and nose rings. Knibbe et al., for example, used conductive fabric electrodes that could be built into a flexible form factor~\cite{knibbeSkillSleevesDesigningElectrode2021}, but still required gel, which often needs periodic replacement. Recent smart-textile work has explored carbon-based dry electrodes integrated into fabric to address this issue~\cite{garnierNovelFunctionalElectrical2024}. For VNS, reliable electrical contact is particularly important: some participants initially reported tingling discomfort, which was resolved after adding more electrode gel. Future work should investigate whether gel-free, flexible electrodes can provide sufficient comfort and contact quality for longitudinal tcVNS deployment studies.

\subsubsection{Real-World Deployment}

In deployment, AppetiteCheck would not require continuous stimulation. Our study tested momentary stimulation during eating, suggesting a use pattern in which the device remains worn but stimulation is active only around eating episodes or high-risk contexts for overeating. Since our in-lab study does not establish how long the behavioral effects last, long-term use may require intermittent refresher sessions. Primary sessions could be user-initiated before or during moments when a user expects difficulty regulating appetite, or triggered through eating episode detection, about-to-eat signals \cite{rahmanPredictingAbouttoEat2016}, or contextual information from wearable sensors. User initiation preserves agency but requires awareness and motivation; closed-loop stimulation may better remain in the background but requires conservative limits on timing, intensity, and frequency, as well as clear user override. Intensity could be bounded by user-calibrated comfort thresholds, session lockouts, and fixed upper limits.

% Truvaga's sinusoidal burst stimulation pattern is patented, so we aim to explore standard square waves or alternative stimulation patterns and validate these, in addition to safety testing. Our pilot testing used square waves before adopting the Truvaga device, so we expect effects to remain consistent.

The appropriate target populations for tcVNS also require careful study. People seeking support for weight management or obesity-related overeating may be a promising target, but a system that can reduce appetite or slow intake could be inappropriate or harmful for people with current or past eating disorders, undernutrition, or medical conditions in which reduced intake is undesirable. Clinical or at-risk populations may also require different stimulation parameters, monitoring, and clinician involvement than the healthy participants in our feasibility study. In this work, we primarily investigated the interaction and initial feasibility of a low-attention physiological intervention. Clinical efficacy in obesity or other populations requires further study with additional safeguards in future work.

\subsection{Limitations} \label{limitations}

\subsubsection{Altered Behavior Due to Experimental Conditions}

During the study, participants' behavior did not necessarily match what they would ordinarily do when snacking. Some participants indicated that they do not usually watch any videos while snacking. Some participants noted they weren't particularly interested in any of the snacks offered, so they chose the healthy option (raisins) or the savory option (chips). Some participants wished to mix and match snacks because they felt they were eating too much of the same thing.

The necessity for the experimenter to hold the stimulator in place also led some participants to slow or stop eating while receiving stimulation. A few participants remarked that the pressure on their neck was uncomfortable, so they ended up eating less. A wearable version of the device could reduce this problem, as the additional pressure arose from the need to maintain contact quality with a handheld device. In the sham stimulation condition, two participants also appeared to accidentally receive brachial plexus stimulation, a collection of arm nerves on the shoulder, due to the stimulation location. As a result, they observed unexpected effects during the sham condition, such as tingling on their fingers.

Although participants were not told which condition was tcVNS or how stimulation was expected to affect eating, expectation or response bias cannot be fully ruled out. The sham condition followed prior work by applying stimulation to a different body location~\cite{rodenkirch_rapid_2022}, which controlled for device contact and tactile stimulation but could not produce a identical sensory experience. Participants rated tcVNS as affecting their eating more than sham stimulation, despite reporting no significant difference in perceived strength or frequency. Because the primary behavioral outcomes were measured by food weight and eating duration rather than self-report, this perception difference does not fully explain the findings, but it remains a limitation of the sham design.

The lab setting and the direct delivery of the intervention by an experimenter may have affected heart rate and HRV, as participants are typically more stressed in the lab setting. Past studies have observed similar findings, where the effect of the intervention on HRV was not significant when evaluating behavioral interventions in a controlled lab setting~\cite{zhaoAffectiveTouchImmediate2023a,gemiciogluBreathePulsePeripheralGuided2024b}. These effects may differ in more naturalistic conditions, and more longitudinal data collection would significantly improve HRV data quality.

% Worth acknowledging that we don't systematically ask participants to profile the sensation, so this is just based on what participants felt like sharing. Also entirely possible that the pressure during manual application could be masking other sensations (or deprioritizing their reporting).

\subsubsection{Snacking and Meal Pacing}

Although we expected participants to be normally distributed within the amount of snacks we provided, the actual amount eaten was far higher, with all the snacks being finished in half the sessions. We suspect that some participants exhibited a ``clean plate'' mentality and simply tried to finish whatever food was offered. A participant remarked that they would have kept eating even if they had been offered a 1kg jar of M\&Ms. Several participants also noted that they had skipped a meal due to the study time or the 3-hour pre-study fasting requirement, making them feel hungrier than they would be in a typical snacking episode.

The nearly binary divide in eating behavior complicated some of the analyses. For example, in food consumption, participants who finished early by eating all the foods in both conditions showed no change, even though their durations were different. Thus, the eating rate was the only behavioral metric capable of accounting for both groups of participants. In a deployment study, such effects could be eliminated due to normal meal and snacking behaviors by participants, but the diversity in both eating habits and foods would greatly increase complexity.

Snacking while distracted also induces only a limited range of eating behaviors, in which the eating pace is deliberately slowed by engaging in a video. As a result, this task may have made it easier for stimulation effects to occur during the eating episode. Moreover, hunger and appetite often have a weaker effect on the amount of food consumed in uncontrolled eating situations where food is eaten rapidly. Realistically, users may also be in a rush and eat very quickly, in which case it could be necessary to activate stimulation before eating. The temporality of tcVNS must be further studied with a broader range of eating behaviors to ensure efficacy.

\subsubsection{Generalizing from Healthy Participants}

Participants in the AppetiteCheck study were healthy and did not have any eating disorders. Their body weight and BMI were not collected to maintain privacy, so whether they are overweight or obese is unknown. In addition, some participants remarked that they would not be interested in using the device, as they had healthy eating habits and did not see a need for any tool to manage their eating behavior. Although past studies of implanted vagus nerve stimulation demonstrated greater effectiveness in patients with obesity~\cite{pardo_weight_2007}, more work with clinical populations is needed to further evaluate AppetiteCheck's effectiveness.

% Although partic

% \subsection{Future Work}

% \subsubsection{Wearable VNS}

% Form factor recommended by 2-3 participants, addresses others' concerns.

% \subsubsection{Closed-Loop Eating Management}

% 
% \cite{rahman_predicting} paper on predicting eating events in advance, flexible temporality

\section{Conclusion}
This paper introduced AppetiteCheck, a handheld device that can be employed in ubiquitous, mobile interventions for managing eating behavior. What is unique about AppetiteCheck is that it is not an overt intervention, and, instead, provides a low-effort intervention based on non-invasive vagus nerve stimulation (tcVNS).

In our in-lab study with 24 participants, food intake was 9.6\% lower and eating rate was 23.6\% slower during distracted snacking under tcVNS compared to sham stimulation. Participants reported the same post-snacking satiety in both conditions and described the stimulation as subtle and barely noticeable. Heart rate was lower during tcVNS, although HRV measures did not differ by condition. These results provide initial evidence that non-invasive vagus nerve stimulation may function as a momentary, low-attention intervention during eating, motivating future ubiquitous systems that pair eating-related sensing with physiological actuation. Together, these findings motivate future work on integrating physiological actuation into wearable, self-administered, and closed-loop eating interventions while preserving comfort, safety, and user control.

\begingroup
\makeatletter
\let\@secfont\@titlecasesecfont
\makeatother
\begin{acks}
This research was supported by the National Science Foundation grants \#2212351 and \#2212352. Any opinions, findings, conclusions, or recommendations expressed in this
material are those of the authors and do not necessarily reflect the
views of the funding body.
\end{acks}
\endgroup

\bibliographystyle{ACM-Reference-Format}
\balance
\bibliography{ref_main}

\clearpage
\appendix

\section*{Appendix A: Qualitative Codebook and Normalization}

To support transparency, this appendix documents the final qualitative codebook used to annotate participant free-text responses, as well as the normalization rules applied to reduce lexical variability.

\subsection*{Sensation}
Final categories included \textit{vibration, tingling, buzzing, hum, pulsing, soreness}. These categories were chosen to preserve perceptually distinct qualities commonly referenced in haptic descriptions and were sourced directly from responses. Semantically overlapping terms (e.g., tingly/tingling) were normalized to a single category.

\subsection*{Intensity}
Final categories included: \textit{subtle} and \textit{unnoticeable}. Intensity-related terms describing low-magnitude sensations (e.g., light, gentle, small, slight, dull) were collapsed into \textit{subtle}, while reports indicating absence of perceptual awareness were coded as \textit{unnoticeable}.

\subsection*{Temporal Characteristics}
Final categories included: \textit{noticeable initially, intermittent, pulsing, fading, smooth}. Temporal categories were defined to distinguish initial salience, variability in presence, rhythmic modulation, and changes in sensation over time. Pulsing was retained as distinct from intermittent, as it reflects structured rhythmic modulation rather than variability in awareness.

% \newpage

\end{document}